\documentclass{ifacconf}
\usepackage{graphicx}      
\usepackage[round]{natbib} 
\usepackage{amsmath}
\usepackage{epstopdf}
\usepackage{graphics} 
\usepackage{epsfig} 
\usepackage{mathptmx} 
\usepackage{times} 
\usepackage{amsmath} 
\usepackage{amssymb}  
\usepackage{enumerate}
\usepackage{mathrsfs}
\usepackage{multirow}
\usepackage{subfigure}
\usepackage{latexsym}
\usepackage{bm}
\usepackage{dsfont}
\usepackage{graphicx}
\usepackage{algpseudocode}
\usepackage{algorithm}

\newtheorem{proposition}{Proposition}

\newtheorem{theorem}{Theorem}

\newtheorem{assumption}{Assumption}

\usepackage{color}

\begin{document}
\begin{frontmatter}
\title{A hierarchical MPC scheme for interconnected systems}
\author[First]{M. Farina}
\author[First]{X. Zhang}
\author[First]{R. Scattolini}
\address[First]{Dipartimento di Elettronica, Informazione e Bioingegneria, Politecnico di Milano, Milan 20133, Italy (e-mail:~marcello.farina@polimi.it).}
\begin{abstract}                
This paper describes a hierarchical control scheme for interconnected systems. The higher layer of the control structure is designed with robust Model Predictive Control (MPC) based on a reduced order dynamic model of the overall system and is aimed at optimizing long-term performance, while at the lower layer local regulators acting at a higher frequency are designed for the full order models of the subsystems to refine the control action. A simulation experiment concerning the control of the temperature inside a building is reported to witness the potentialities of the proposed approach.
\end{abstract}

\begin{keyword}
Hierarchical MPC, robust MPC, multivariable systems.
\end{keyword}

\end{frontmatter}
%
\section{Introduction and main idea} \label{Sec:Intro} 
\begin{figure*}[ht]
\center
\includegraphics[width=0.7\linewidth]{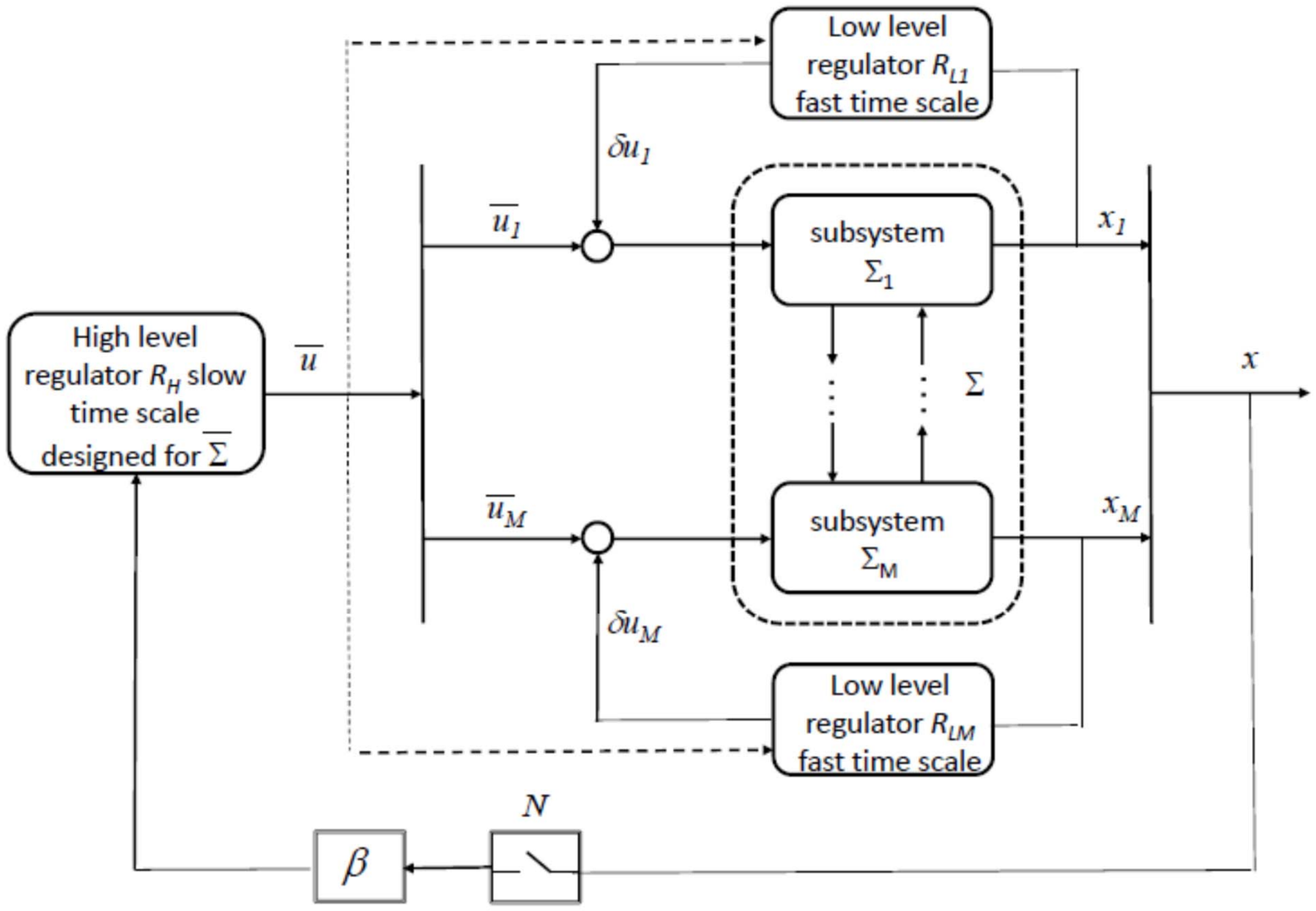}
\caption{Overall control scheme.}
\label{fig:GS}
\end{figure*}

Physical and cyber-physical systems are becoming  more and more complex, large-scale, and heterogeneous due to the growing opportunities provided by information technology in terms of computing power, transmission of information, and networking capabilities. As a consequence, also the management and control of these systems represents a problem of increasing difficulty and requires innovative solutions. A classical approach to deal with this challenge consists of resorting to hierarchical control structures, where at the higher layer of the hierarchy simplified models are used to predict and control the long term behavior of the overall system, while at the lower layer local control actions are designed to compensate for model inaccuracies, disturbances, or parametric variations. Along this line, many hierarchical control methods have been described in the past, see e.g. \cite{Adetola}, \cite{Kadam} in the context of Real Time Optimization (RTO), or \cite{Amrit,Grune,Diehl} in the emerging area of economic MPC.\\
In view of the potentialities of multilayer control structures, this paper describes a novel approach to the design of a hierarchical control structure for large scale systems composed by interconnected subsystems. The scheme of the proposed solution is sketched in Figure~\ref{fig:GS}: the system under control $\Sigma$ is composed of $M$ interconnected subsystems $\Sigma_{1},...,\Sigma_M$. A reduced order model $\bar{\Sigma}_i$, $i=1,...,M$ is computed for each subsystem, and the overall reduced order model $\bar{\Sigma}$ is obtained; typically  $\bar{\Sigma}_i$ and $\bar{\Sigma}$ represent low-frequency approximations of the corresponding systems. At a slow sampling rate, a centralized MPC regulator $R_{\rm\scriptscriptstyle H}$ is designed for $\bar{\Sigma}$ to consider the long-term behavior of the controlled system and to compute the control variables $\bar{u}_i$, $i=1,...,M$. Then, local regulators $R_{Li}$, $i=1,...,M$, working at a faster time scale, are designed for each subsystem $\Sigma_{_{i}}$: their scope is to compute the control contributions $\delta u_i$ compensating for the inaccuracies in the high layer design due to the mismatch between $\Sigma$ and $\bar{\Sigma}$. This structure has already been studied in \cite{PicassoZhangScat} where, however, only independent systems $\Sigma_{i}$ with joint output constraints were considered. The advantage of the approach here proposed is twofold: first, at the slower time scale the optimization problem underlying the MPC solution is of reduced dimension and can minimize a global cost function over a long horizon with a limited computational cost; second, also the local regulators designed for the local subsystems involve the solution to optimization problems whose complexity only depends on the order of the local submodels.\\
The paper is organized as follows. In Section 2 the models considered at the two layers of the control structure are introduced. Section 3 describes the MPC algorithms adopted at the two layers, while Section 4 is presents the main feasibility and convergence results as well as a summary of the main steps to be performed in the algorithm implementation. Section 5 describes a simulation example, while in Section 6 some conclusions are drawn. The proofs of the main results are reported in the Appendix.\\
\textbf{Notation:} for a given a set of variables $z_{i}\in{\mathbb{R}}^{q_{i}}$,
$i=1,2,\dots,M$, we define the vector whose vector-components are $z_{i}$ in the following compact form: $(z_{1}, z_{2}, \cdots, z_{M})=[\,z_{1}^T\ z_{2}^T\ \cdots\ z_{M}^T\,]^T\in{\mathbb{R}}^{q}$, where $q= \sum_{i=1}^{M}q_{i}$.
The symbols $\oplus$/$\ominus$ denote the Minkowski sum/difference. We denote with $\|\cdot\|$ the Euclidean norm.
Finally, a ball with radius $\rho_{\varepsilon_{i}}$ and centered at $\bar{x}$ in the ${\mathbb{R}}^{dim}$
space is defined as follows
\[
\mathcal{B}_{\rho_{\varepsilon_{i}}}(\bar{x}):=\{x\in{\mathbb{R}}^{dim}:||x-\bar{x}||\leq\rho_{\varepsilon_{i}}\}
\]
\section{Models for the two-layer control scheme}
\label{Sec:linear subsystem}
In this section we present the model of the complex system under study and the simplified one used for high-level control.
\subsection{Large-scale system model}
In line with \cite{LunzeBook92}, we assume that the overall system $\Sigma$ is
composed by $M$ discrete-time, linear, interacting subsystems described by
\begin{equation}
\Sigma_{i}:\ \left\{ \begin{array}{lcl}
x_{i}(h+1)&=&A_{\rm\scriptscriptstyle L}^{ii}x_{i}(h)+B_{\rm\scriptscriptstyle L}^{ii}u_{i}(h)+E_{\rm\scriptscriptstyle L}^{i}s_{i}(h)\\
z_{i}(h)&=&C_{\rm\scriptscriptstyle L}^{zi}x_{i}(h),
\end{array}\right.\label{Eqn:LL}
\end{equation}
$i=1,2,\dots,M$, where $x_{i}\subseteq{\mathbb{R}}^{n_{i}}$, $u_{i}\in{\mathcal{U}}_{i}\subseteq{\mathbb{R}}^{m_{i}}$
are the state and input vectors, and where the interconnections among the $\Sigma_{i}^{'}s$ are represented by the coupling input and output vectors $s_{i}\in{\mathbb{R}}^{p_{si}}$
and $z_{i}\in{\mathbb{R}}^{p_{zi}}$ linked through the following expression
\begin{equation}
s_{i}(h)=\sum_{j=1}^{M}L_{ij}z_{j}(h)\label{Eqn:si}
\end{equation}
with $L_{ii}=0$, $i=1,...,M$. The sets ${\mathcal{U}}_i$ are closed and convex sets containing the origin.\\
\\
%
Collecting all the subsystems \eqref{Eqn:LL}, the overall dynamical model of $\Sigma$ is
\begin{equation}
\Sigma:x(h+1)=A_{\rm\scriptscriptstyle L}x(h)+B_{\rm\scriptscriptstyle L}u(h)\label{Eqn:C full model}
\end{equation}
where $x=(x_{1},\ldots,x_{M})\in\mathbb{R}^{n}$, $n=\sum_{i=1}^{M}n_{i}$,
$u=(u_{1},\ldots,u_{M})\in\mathbb{R}^{m}$, $m=\sum_{i=1}^{M}m_{i}$. The diagonal blocks of $A_L$
are state transition matrices $A_{\rm\scriptscriptstyle L}^{ii}$, whereas
the coupling terms among the subsystems correspond to the non-diagonal
blocks of $A_{\rm\scriptscriptstyle L}$, i.e., $A_{\rm\scriptscriptstyle L}^{ij}=E_{\rm\scriptscriptstyle L}^{i}L_{ij}C_{\rm\scriptscriptstyle L}^{zj}$,
with $j\neq i$. The collective input matrix is $B_{\rm\scriptscriptstyle L}=$diag$(B_{\rm\scriptscriptstyle L}^{11},...,\,B_{\rm\scriptscriptstyle L}^{MM})$.
We define ${\mathcal{U}}=\prod_{i=1}^{M}{\mathcal{U}}_{i}\subseteq\mathbb{R}^{m}$,
which is convex by the convexity of $\mathcal{U}_{i}$.\\
Concerning systems \eqref{Eqn:LL} and \eqref{Eqn:C full model}, the
following standing assumption is introduced:
\begin{assumption}\label{eq:Assump_LL}\hfill
\begin{enumerate}
\item \label{eq:Assump_LL_statefeedback} the state $x_{i}$ is measurable, for each $i=1,\dots,M$;
\item \label{eq:Assump_LL_Schur} $A_{\rm\scriptscriptstyle L}$ is Schur stable;
\item \label{eq:Assump_LL_Reach} the pair $(A_{\rm\scriptscriptstyle L}^{ii},\,B_{\rm\scriptscriptstyle L}^{ii})$ is
reachable, for each $i=1,\dots,M$.\hfill$\square$
\end{enumerate}\end{assumption}
Note that Assumption \ref{eq:Assump_LL}.\ref{eq:Assump_LL_statefeedback} can be relaxed, using suitable observers - e.g., distributed ones - and accounting for state estimation error in a rigorous way.
\subsection{Reduced order models}
Associated with each subsystem $\Sigma_{i},\,i=1,...,\,M$, consider a reduced order model $\bar{\Sigma}_{i},\,i=1,...,\,M$, with state
$\bar{x}_{i}\in\mathbb{R}^{\bar{n}_{i}}$, $\bar{n}_i \leq n_i$, and input $\bar{u}_{i}\in{\mathcal{\bar{U}}}_{i}\subseteq{\mathcal{U}}_{i}$.
%
%
%
In a collective form, these systems $\bar{\Sigma}_i$
define the overall reduced order model
\begin{equation}
\bar{\Sigma}:\ \begin{array}{c}
\bar{x}(h+1)=A_{\rm\scriptscriptstyle H}\bar{x}(h)+B_{\rm\scriptscriptstyle H}\bar{u}(h)\end{array}\label{Eqn:C HL model}
\end{equation}
where $\bar{x}=(\bar{x}_{1},\ldots,\bar{x}_{M})\in\mathbb{R}^{\bar{n}}$,
$\bar{n}=\sum_{i=1}^{M}\bar{n}_{i}$, and $\bar{u}=(\bar{u}_{1},\ldots,\bar{u}_{M})\in{\mathcal{\bar{U}}}=\prod_{i=1}^{M}{\mathcal{\bar{U}}}_{i}\subseteq\mathbb{R}^{m}$.\\
The reduced order models $\bar{\Sigma}_i$ can be defined according to different criteria. In any case, it is required that the stability properties of system $\Sigma$ are inherited by $\bar{\Sigma}$. Moreover, it is assumed that, for each subsystem $i=1,\dots,M$, there exists a state projection
$\beta_i:\mathbb{R}^{{n}_{i}}\rightarrow \mathbb{R}^{\bar{n}_{i}}$, $i=1,...,M$, that allows to establish a connection between the states $x_i(h)$ of the original models and the states of the reduced models $\bar{x}_i(h)$. Collectively, we define $\beta=\text{diag}(\beta_1,...,\beta_M)$. In principle, the ideal case would be to verify $\bar{x}(h)=\beta x(h)$ for all $h\geq 0$. However, due to model reduction approximations, this ideal assumption must be relaxed; instead, we just ask that $\bar{x}(h)=\beta x(h)$ at least in steady-state conditions. Overall, we require the following standing assumption to be satisfied.
\begin{assumption}\label{Assump:3}\hfill
\begin{enumerate}
\item \label{Assump:3_0} $A_{\rm\scriptscriptstyle H}$ is Schur stable;
\item \label{Assump:3_1} $\beta_i$ is full rank, for each $i=1,\dots,M$;
\item \label{Assump:3_2} letting $\hat{G}_{\rm\scriptscriptstyle L}(z)=\beta (zI-A_{\rm\scriptscriptstyle L})^{-1}B_{\rm\scriptscriptstyle L}$ and $G_{\rm\scriptscriptstyle H}(z)=(zI-A_{\rm\scriptscriptstyle H})^{-1}B_{\rm\scriptscriptstyle H}$, it holds that $\hat{G}_{\rm\scriptscriptstyle L}(1)={G}_{\rm\scriptscriptstyle H}(1)$.\hfill$\square$
\end{enumerate}\end{assumption}
An algorithm to compute the projections $\beta_i$ and the matrices of $\bar{\Sigma}$ can be devised along the lines of~\cite{PicassoZhangScat}.
\section{Design of the hierarchical control structure}
In this section the regulators at the two layers of the hierarchical control structure are designed.

\subsection{Design of the high level regulator}
\label{Sec:HLregulator}
The high level regulator, designed to work at a lower frequency, is based on the reduced order model \eqref{Eqn:C HL model} sampled
with period $N_{\rm\scriptscriptstyle L}$ under the assumption that, $\forall\,k\in{\mathbb{N}}$,
the $\bar{u}_{i}'s$ are held constant over the interval $h\in[kN_{\rm\scriptscriptstyle L},kN_{\rm\scriptscriptstyle L}+N_{\rm\scriptscriptstyle L}-1]$.
Denoting by $\bar{u}_{i}^{{[N_{\rm\scriptscriptstyle L}]}}(k)$ these constant
values and by $\bar{u}^{{[N_{\rm\scriptscriptstyle L}]}}(k)$ the overall
input vector, the reduced order model in the slow timescale is
\begin{equation}
\bar{\Sigma}^{{[N_{\rm\scriptscriptstyle L}]}}:\ \begin{array}{l}
\bar{x}^{{[N_{\rm\scriptscriptstyle L}]}}(k+1)=A_{{\rm {\rm\scriptscriptstyle H}}}^{N_{\rm\scriptscriptstyle L}}\bar{x}^{{[N_{\rm\scriptscriptstyle L}]}}(k)+B_{{\rm {\rm\scriptscriptstyle H}}}^{{[N_{\rm\scriptscriptstyle L}]}}\bar{u}^{{[N_{\rm\scriptscriptstyle L}]}}(k)\end{array}\label{Eqn:C HLS model}
\end{equation}
where $B_{{\rm {\rm\scriptscriptstyle H}}}^{{[N_{\rm\scriptscriptstyle L}]}}=\sum_{j=0}^{N_{\rm\scriptscriptstyle L}-1}A_{{\rm {\rm\scriptscriptstyle H}}}^{j}B_{{\rm {\rm\scriptscriptstyle H}}}$.
In order to feedback a value of $\bar{x}^{[N_{\rm\scriptscriptstyle L}]}$ related to the real
state $x$ of the system, the projected value $\beta_{i}x_{i}(kN_{\rm\scriptscriptstyle L})$ must be used, so that the reset
\begin{equation}
\bar{x}_{i}^{[N_{\rm\scriptscriptstyle L}]}(k)=\beta_{i}x_{i}(kN_{\rm\scriptscriptstyle L})\label{Eqn:betai}
\end{equation}

must be applied. In collective form \eqref{Eqn:betai} becomes
\begin{equation}\bar{x}^{[N_{\rm\scriptscriptstyle L}]}(k)=\beta x(kN_{\rm\scriptscriptstyle L})\label{eq:reset_bar}\end{equation}
The reset \eqref{Eqn:betai} at time $k$
may force $\bar{x}^{[N_{\rm\scriptscriptstyle L}]}(k+1)$ to assume a value different from
the one computed based on the dynamics of \eqref{Eqn:C HLS model}
and the applied input $\bar{u}^{[N_{\rm\scriptscriptstyle L}]}(k)$. This discrepancy, due to the model reduction error and to the actions of the low level controllers,
is accounted for by including in \eqref{Eqn:C HLS model} an additive disturbance $\bar{w}(k)$, i.e.,
\begin{equation}
\bar{\Sigma}_{w}^{{[N_{\rm\scriptscriptstyle L}]}}:\ \begin{array}{l}
\bar{x}^{{[N_{\rm\scriptscriptstyle L}]}}(k+1)=A_{{\rm {\rm\scriptscriptstyle H}}}^{N_{\rm\scriptscriptstyle L}}\bar{x}^{{[N_{\rm\scriptscriptstyle L}]}}(k)+B_{{\rm {\rm\scriptscriptstyle H}}}^{{[N_{\rm\scriptscriptstyle L}]}}\bar{u}^{{[N_{\rm\scriptscriptstyle L}]}}(k)+\bar{w}(k)\end{array}\label{Eqn:C HLS model1}
\end{equation}
The size of $\bar{w}(k)$ depends on the action of the low level regulators and its presence requires to resort to a robust MPC method, which is here designed assuming that $\bar{w}(k)\in \mathcal{W}$, where $\mathcal{W}$ is a compact set containing the origin. The characteristics of $\mathcal{W}$ will be defined in the following once the low level regulators have been specified (see Section 4).\\
The robust MPC algorithm is based on the scheme proposed in \cite{Mayne2005219}. To this end, we first need to define the ``unperturbed" prediction model
\begin{equation}
\bar{\Sigma}_{w}^{{[N_{\rm\scriptscriptstyle L}]},o}:\ \begin{array}{l}
\bar{x}^{{[N_{\rm\scriptscriptstyle L}]},o}(k+1)=A_{{\rm {\rm\scriptscriptstyle H}}}^{N_{\rm\scriptscriptstyle L}}\bar{x}^{{[N_{\rm\scriptscriptstyle L}]},o}(k)+B_{{\rm {\rm\scriptscriptstyle H}}}^{{[N_{\rm\scriptscriptstyle L}]}}\bar{u}^{{[N_{\rm\scriptscriptstyle L}]},o}(k)\end{array}\label{Eqn:C HLS model_unp}
\end{equation}
and the control gain matrix $\bar{K}_{\rm\scriptscriptstyle H}$ such that, at the same time
\begin{itemize}
\item $F_{\rm\scriptscriptstyle H}=A_{{\rm {\rm\scriptscriptstyle H}}}^{N_{\rm\scriptscriptstyle L}}+B_{{\rm {\rm\scriptscriptstyle H}}}^{{[N_{\rm\scriptscriptstyle L}]}}\bar{K}_{\rm\scriptscriptstyle H}$ is Schur stable.
\item $F_{\rm\scriptscriptstyle L}^{[N_{\rm\scriptscriptstyle L}]}=A_{\rm\scriptscriptstyle L}^{N_{\rm\scriptscriptstyle L}}+B_{\rm\scriptscriptstyle L}^{[N_{\rm\scriptscriptstyle L}]}\bar{K}_{\rm\scriptscriptstyle H}\beta$ is Schur stable, where $B_{\rm\scriptscriptstyle L}^{[N_{\rm\scriptscriptstyle L}]}=\sum_{j=0}^{N_{\rm\scriptscriptstyle L}-1}A_{{\rm {\rm\scriptscriptstyle L}}}^{j}B_{\rm\scriptscriptstyle L}$.
\end{itemize}
We define $\bar{e}(k)=\bar{x}^{{[N_{\rm\scriptscriptstyle L}]},o}(k)-\bar{x}^{{[N_{\rm\scriptscriptstyle L}]},o}(k)$ and we let $\mathcal{Z}$ be a robust positively invariant (RPI) set - minimal, if possible - for the autonomous but perturbed system
\begin{equation}
\bar{\Sigma}_{w}^{{[N_{\rm\scriptscriptstyle L}]},e}:\ \begin{array}{l}
\bar{e}(k+1)=F_{\rm\scriptscriptstyle H}\bar{e}(k)+\bar{w}(k)\end{array}\label{Eqn:C HLS model_err}
\end{equation}
Denoting by $\overrightarrow{\bar{u}^{{[N_{\rm\scriptscriptstyle L}]},o}}{(t:t+N_{\rm\scriptscriptstyle H}-1)}$ the sequence $\bar{u}^{{[N_{\rm\scriptscriptstyle L}]},o}(t)$, $\dots$, $\bar{u}^{{[N_{\rm\scriptscriptstyle L}]},o}(t+N_{\rm\scriptscriptstyle H}-1)$, at each slow time-step $t$ the following optimization problem is solved:
\begin{equation}
\begin{array}{l}
\min_{\bar{x}^{{[N_{\rm\scriptscriptstyle L}]},o}(t),\overrightarrow{\bar{u}^{{[N_{\rm\scriptscriptstyle L}]},o}}{(t:t+N_{\rm\scriptscriptstyle H}-1)}}{J_{{\rm\scriptscriptstyle H}}\big(\bar{x}^{{[N_{\rm\scriptscriptstyle L}]},o}(t),\overrightarrow{\bar{u}^{{[N_{\rm\scriptscriptstyle L}]},o}}{(t:t+N_{\rm\scriptscriptstyle H}-1)})}\\
\mbox{subject to:}\\
\bullet\,\,\mbox{the unperturbed model dynamics}~\eqref{Eqn:C HLS model_unp}\\
\bullet\,\,\mbox{the initial constraint}~\beta x(tN_{\rm\scriptscriptstyle L})-\bar{x}^{{[N_{\rm\scriptscriptstyle L}]},o}(t)\in \mathcal{Z}\\
\bullet\,\,\mbox{the terminal constraint}~\bar{x}^{{[N_{\rm\scriptscriptstyle L}]},o}(t+N_{\rm\scriptscriptstyle H})\in \bar{\mathcal{X}}_F\\
\bullet\,\, \bar{u}^{{[N_{\rm\scriptscriptstyle L}]},o}(k)\in{\bar{\mathcal{U}}\ominus \bar{K}_{\rm\scriptscriptstyle H}\mathcal{Z}},\,k=t,\dots,t+N_{\rm\scriptscriptstyle H}-1,
\end{array}\label{Eqn:HLoptimiz_1}
\end{equation}
where
\begin{equation}
J_{{\rm\scriptscriptstyle H}}=\sum_{k=t}^{t+N_{\rm\scriptscriptstyle H}-1}\|\bar{x}^{{[N_{\rm\scriptscriptstyle L}]},o}(k)\|_{Q_{\rm\scriptscriptstyle H}}^{2}+\|\bar{u}^{{[N_{\rm\scriptscriptstyle L}]},o}(k)\|_{R_{\rm\scriptscriptstyle H}}^{2}+\|\bar{x}^{{[N_{\rm\scriptscriptstyle L}]},o}(t+N_{\rm\scriptscriptstyle H})\|_{P_{\rm\scriptscriptstyle H}}^{2}
\label{Eqn:LL_cost}
\end{equation}
and $\bar{\mathcal{X}}_F$ is a positively invariant terminal set for the unperturbed system~\eqref{Eqn:C HLS model_unp} controlled with the stabilizing auxiliary control law  $\bar{u}^{{[N_{\rm\scriptscriptstyle L}]},o}(k)=\bar{K}_{\rm\scriptscriptstyle H} \bar{x}^{{[N_{\rm\scriptscriptstyle L}]},o}(k)$, with $\bar{K}_{\rm\scriptscriptstyle H}\bar{\mathcal{X}}_F\subseteq {\bar{\mathcal{U}}\ominus \bar{K}_{\rm\scriptscriptstyle H}\mathcal{Z}}$. Note that it is implicitly assumed that ${\bar{\mathcal{U}}\supset \bar{K}_{\rm\scriptscriptstyle H}\mathcal{Z}}$: this can always made possible by reducing $\bar{K}_{\rm\scriptscriptstyle H}$ and set $\mathcal{W}$ and - in turn - $\mathcal{Z}$; as it will be discuss in the following, the latter can be reduced, for example, by increasing $N_{\rm\scriptscriptstyle L}$. The positive definite weighting matrices ${Q_{\rm\scriptscriptstyle H}}$, ${R_{\rm\scriptscriptstyle H}}$ are free design parameters, while  ${P_{\rm\scriptscriptstyle H}}$ is computed as the solution to the Lyapunov equation
\begin{equation}
F_{\rm\scriptscriptstyle H}^T P_{\rm\scriptscriptstyle H} F_{\rm\scriptscriptstyle H}-P_{\rm\scriptscriptstyle H}=-(Q_{\rm\scriptscriptstyle H}+\bar{K}_{\rm\scriptscriptstyle H}^TR_{\rm\scriptscriptstyle H} \bar{K}_{\rm\scriptscriptstyle H})
\label{eqn:HL_Lyap}
\end{equation}
Letting $\bar{x}^{{[N_{\rm\scriptscriptstyle L}]},o}(t|t),\overrightarrow{\bar{u}^{{[N_{\rm\scriptscriptstyle L}]},o}}{(t:t+N_{\rm\scriptscriptstyle H}-1|t)}$ be the solution to the optimization problem~\eqref{Eqn:HLoptimiz_1}, the control variable applied at time $t$ is defined as
\begin{equation}
\bar{u}^{{[N_{\rm\scriptscriptstyle L}]}}(t)=\bar{u}^{{[N_{\rm\scriptscriptstyle L}]},o}(t|t)+\bar{K}_{\rm\scriptscriptstyle H}(\beta x(tN_{\rm\scriptscriptstyle L})-\bar{x}^{{[N_{\rm\scriptscriptstyle L}]},o}(t|t))
\label{eqn:HL_control_input}
\end{equation}
%
%
\subsection{Design of the low level regulators}
Recall that (see again Figure~\ref{fig:GS}) the overall control action has components generated by both the high-level and the low-level controllers, i.e.,
\begin{equation}
u_i(h)=\bar{u}_{i}^{[N_{\rm\scriptscriptstyle L}]}(\lfloor h/N_{\rm\scriptscriptstyle L} \rfloor)+\delta u_i(h)\label{contrcompl}
\end{equation}
Indeed, the low level regulators are in charge of computing the local control corrections $\delta u_{i}\in{\mathcal{U}}_{i}\ominus{\bar{\mathcal{U}}}_{i}$ compensating for the effect of the model inaccuracies at the high level expressed by the term $\bar{w}(k)$ in \eqref{Eqn:C HLS model1}.
To this end, first define the auxiliary system $\hat{\Sigma}_{i}$ given by
\begin{equation}
\hat{\Sigma}_{i}:\ \left\{ \begin{array}{lcl}
\hat{x}_{i}(h+1)&=&A_{\rm\scriptscriptstyle L}^{ii}\hat{x}_{i}(h)+B_{\rm\scriptscriptstyle L}^{ii}\bar{u}_{i}^{{[N_{\rm\scriptscriptstyle L}]}}(\lfloor h/N_{\rm\scriptscriptstyle L}\rfloor)+E_{\rm\scriptscriptstyle L}^{i}\hat{s}_{i}(h)\\
\hat{s}_{i}(h)&=&\sum_{j=1}^{M}L_{ij}\hat{z}_{j}(h)\\
\hat{z}_{i}(h)&=&C_{\rm\scriptscriptstyle L}^{zi}\hat{x}_{i}(h)\\
\hat{x}_{i}(kN_{\rm\scriptscriptstyle L})&=&x_{i}(kN_{\rm\scriptscriptstyle L})
\end{array}\right.\label{Eqn:LL sdyn}
\end{equation}
Note that $\hat{\Sigma}_{i}$ can be simulated in a centralized way in the time interval $[kN_{\rm\scriptscriptstyle L},\,kN_{\rm\scriptscriptstyle L}+N_{\rm\scriptscriptstyle L})$
once the high level controller has computed $\bar{u}_{i}^{{[N_{\rm\scriptscriptstyle L}]}}(k)$.

Also denote by $\varDelta\Sigma_{i}$ the model given by the difference
of the system \eqref{Eqn:LL}, with \eqref{Eqn:si}, \eqref{contrcompl}, and \eqref{Eqn:LL sdyn} in the
form
\begin{equation}
\varDelta\Sigma_{i}:\ \left\{ \begin{array}{lcl}
\delta x_{i}(h+1)&=&A_{\rm\scriptscriptstyle L}^{ii}\delta x_{i}(h)+B_{\rm\scriptscriptstyle L}^{ii}\delta u_{i}(h)+E_{\rm\scriptscriptstyle L}^{i}\delta s_{i}(h)\\
\delta{s}_{i}(h)&=&\sum_{j=1}^{M}L_{ij}\delta{z}_{j}(h)\\
\delta z_{i}(h)&=&C_{\rm\scriptscriptstyle L}^{zi}\delta x_{i}(h)\\
\delta x_{i}(kN_{\rm\scriptscriptstyle L})&=&0
\end{array}\right.\label{Eqn:LL ddyn}
\end{equation}
where $\delta x_{i}(h)=x_{i}(h)-\hat{x}_{i}(h)$ , $\delta z_{i}(h)=z_{i}(h)-\hat{z}_{i}(h)$
and $\delta s_{i}(h)=s_{i}(h)-\hat{s}_{i}(h)$.

The difference state $\delta x_{i}$ is available at each time instant
$h$ since $x_{i}$ is measurable and $\hat{x}_{i}$ can be computed
from the available control $\bar{u}_{i}^{{[N_{\rm\scriptscriptstyle L}]}}(\lfloor h/N_{\rm\scriptscriptstyle L}\rfloor)$.
However, the difference dynamical system $\varDelta\Sigma_{i}$ is
not yet useful for decentralized prediction since it depends upon the interconnection variables $\delta s_i(h)$ that, in turn, depend upon the variables $\delta x_j(h)$, $j\neq i$, which are not known in advance in the future prediction horizon. For this reason, we define a decentralized (approximated)
dynamical system $\varDelta\hat{\Sigma}_{i}$ by discarding all interconnection inputs and with input $\delta\hat{u}_i(h)$ (which will be defined as the result of a suitable optimization problem), i.e.,
\begin{equation}
\varDelta\hat{\Sigma}_{i}:\ \left\{
\begin{array}{lcl}
\delta\hat{x}_{i}(h+1)&=&A_{\rm\scriptscriptstyle L}^{ii}\delta\hat{x}_{i}(h)+B_{\rm\scriptscriptstyle L}^{ii}\delta \hat{u}_{i}(h)\\
\delta\hat{x}_{i}(kN_{\rm\scriptscriptstyle L})&=&0
\end{array}\right.\label{Eqn:LL dndyn}
\end{equation}
For all $i=1,\dots,M$, the input $\delta u_i(h)$ to the real model \eqref{Eqn:LL ddyn} is computed based on $\delta \hat{u}_i(h)$, $\delta x_i(h)$, and $\delta \hat{x}_i(h)$ using a standard state-feedback policy, i.e.,
\begin{equation}
\delta u_i(h)=\delta \hat{u}_i(h)+K_i (\delta {x}_i(h)-\delta \hat{x}_i(h))\label{Eqn:LL dndyn_sf_deltau}
\end{equation}
and where $K_i$ is designed in such a way that the matrix $F_{\rm\scriptscriptstyle L}=A_{\rm\scriptscriptstyle L}+B_{\rm\scriptscriptstyle L}K$ is Schur stable, being $K=$diag$(K_1,\dots,K_M)$.\\
Assume now to be at time $h=kN_{\rm\scriptscriptstyle L}$ and to have run the high level controller,
so that both $\bar{u}_{i}^{{[N_{\rm\scriptscriptstyle L}]}}(k)$ and the
predicted value $\bar{x}^{[N_{\rm\scriptscriptstyle L}]}(k+1|k)=A_{\rm\scriptscriptstyle H}^{N_{\rm\scriptscriptstyle L}}\bar{x}^{[N_{\rm\scriptscriptstyle L}]}(k)+B_{\rm\scriptscriptstyle H}^{[N_{\rm\scriptscriptstyle L}]}\bar{u}^{[N_{\rm\scriptscriptstyle L}]}(k)$ are available. Therefore,
in order to remove the effect of the mismatch at the high level represented by $\bar{w}(k)$ in \eqref{Eqn:C HLS model1},
the low level controller working in the interval $[kN_{\rm\scriptscriptstyle L},kN_{\rm\scriptscriptstyle L}+N_{\rm\scriptscriptstyle L}-1]$ should, if possible, aim to fulfill
$$\beta_{i}x_{i}(kN_{\rm\scriptscriptstyle L}+N_{\rm\scriptscriptstyle L})=\bar{x}_{i}^{[N_{\rm\scriptscriptstyle L}]}(k+1|k)$$
or equivalently,
\begin{equation}
\beta_{i}\delta x_{i}(kN_{\rm\scriptscriptstyle L}+N_{\rm\scriptscriptstyle L})=\bar{x}_{i}^{[N_{\rm\scriptscriptstyle L}]}(k+1|k)-\beta_{i}\hat{x}_{i}(kN_{\rm\scriptscriptstyle L}+N_{\rm\scriptscriptstyle L})\label{eq:terminal}
\end{equation}

Since the model used for low-level control design is the decentralized one (i.e., \eqref{Eqn:LL dndyn}), the constraint \eqref{eq:terminal} can only be formulated in an approximated way with reference to its state
$\delta\hat{x}_{i}$ as follows:
\begin{equation}
\beta_{i}\delta\hat{x}_{i}(kN_{\rm\scriptscriptstyle L}+N_{\rm\scriptscriptstyle L})=\bar{x}_{i}^{[N_{\rm\scriptscriptstyle L}]}(k+1|k)-\beta_{i}\hat{x}_{i}(kN_{\rm\scriptscriptstyle L}+N_{\rm\scriptscriptstyle L})\label{eq:terminaln}
\end{equation}
Note however that the fulfillment of \eqref{eq:terminaln} does not
imply that \eqref{eq:terminal} is satisfied due to the neglected
interconnections in \eqref{Eqn:LL dndyn} which make the term $\bar{w}(k)$ in \eqref{Eqn:C HLS model} not identically equal to zero, although it contributes to its reduction.\\
Letting $\overrightarrow{\delta \hat{u}}_i {(kN_{\rm\scriptscriptstyle L}:kN_{\rm\scriptscriptstyle L}+N_{\rm\scriptscriptstyle L}-1)}=(\delta \hat{u}_{i}(kN_{\rm\scriptscriptstyle L}), \dots, \delta \hat{u}_{i}(kN_{\rm\scriptscriptstyle L}+N_{\rm\scriptscriptstyle L}-1))\in({\mathbb{R}}^{m_{i}})^{N_{\rm\scriptscriptstyle L}-1}$,
the low level control action is computed, at time instant $h=kN_{\rm\scriptscriptstyle L}$, based on the solution to the following optimization problem:
\begin{equation}
\begin{array}{l}
\min_{\overrightarrow{\delta \hat{u}}_{i}{(kN_{\rm\scriptscriptstyle L}:kN_{\rm\scriptscriptstyle L}+N_{\rm\scriptscriptstyle L}-1)}} J_{{\rm\scriptscriptstyle L}}\big(\delta\hat{x}_{i}(kN_{\rm\scriptscriptstyle L}),\overrightarrow{\delta \hat{u}}_{i}{(kN_{\rm\scriptscriptstyle L}:kN_{\rm\scriptscriptstyle L}+N_{\rm\scriptscriptstyle L}-1)})\\
\mbox{subject to:}\\
\bullet\,\,\mbox{the dynamics}~\eqref{Eqn:LL dndyn}\\
\bullet\,\,\mbox{the terminal constraint}~\eqref{eq:terminaln}\\
\bullet\,\,\delta \hat{u}_{i}(kN_{\rm\scriptscriptstyle L}+j)\in{\varDelta\hat{\mathcal{U}}}_{i},\,j=0,\dots,N_{\rm\scriptscriptstyle L}-1,
\end{array}\label{Eqn:LLoptimiz_1}
\end{equation}
where
\begin{equation}
J_{{\rm\scriptscriptstyle L}}=\sum_{j=0}^{N_{\rm\scriptscriptstyle L}-1}\|\delta\hat{x}_{i}(kN_{\rm\scriptscriptstyle L}+j)\|_{Q_{i}}^{2}+\|\delta \hat{u}_{i}(kN_{\rm\scriptscriptstyle L}+j)\|_{R_{i}}^{2}
\label{Eqn:LL_cost}
\end{equation}
and where a discussion on how to select the set ${\varDelta\hat{\mathcal{U}}}_{i}$ is deferred to Appendix~\ref{app:rhos}.\\
Finally, at each (fast) time instant, the control component $\delta u_i(kN_{\rm\scriptscriptstyle L}+j)$ is given by $\delta u_i(kN_{\rm\scriptscriptstyle L}+j)=\delta\hat{u}_i(kN_{\rm\scriptscriptstyle L}+j|kN_{\rm\scriptscriptstyle L})+K_i(\delta{x}_i(kN_{\rm\scriptscriptstyle L}+j)-\delta \hat{x}_i(kN_{\rm\scriptscriptstyle L}+j|kN_{\rm\scriptscriptstyle L}))$.
%
%
\section{Properties and algorithm implementation}
\label{sec:rec_feas}
The recursive feasibility and robust convergence properties of the optimization problems stated at the high and low levels are now established. To this end, define
\begin{equation}\kappa(N_{\rm\scriptscriptstyle L})=\|\mathcal{B}(N_{\rm\scriptscriptstyle L})\|\label{Eqn:Gkappa1}\end{equation}
where
$$\mathcal{B}(N_{\rm\scriptscriptstyle L})=\sum_{j=1}^{N_{\rm\scriptscriptstyle L}}A_{\rm\scriptscriptstyle H}^{N_{\rm\scriptscriptstyle L}-j}B_{\rm\scriptscriptstyle H}-\beta\sum_{j=1}^{N_{\rm\scriptscriptstyle L}}A_{\rm\scriptscriptstyle L}^{N_{\rm\scriptscriptstyle L}-j}B_{\rm\scriptscriptstyle L}$$
Also, let ${\mathcal{A}}(N_{\rm\scriptscriptstyle L})=A_{{\rm {\rm\scriptscriptstyle H}}}^{N_{\rm\scriptscriptstyle L}}\beta-\beta A_{{\rm L}}^{N_{\rm\scriptscriptstyle L}}\in{\mathbb{R}}^{\bar{n}\times n}$,
$I_{si}=[0_{\bar{n}_{i},\bar{n}_{1}}\ldots I_{\bar{n}_{i},\bar{n}_{i}}\ldots\\0_{\bar{n}_{i},\bar{n}_{M}}]$,
${\mathcal{R}(N_{\rm\scriptscriptstyle L})}=\big[\,B_{{\rm {\rm\scriptscriptstyle L}}}\ A_{{\rm {\rm\scriptscriptstyle L}}}B_{{\rm {\rm\scriptscriptstyle L}}}\ \cdots\ (A_{{\rm {\rm\scriptscriptstyle L}}})^{N_{\rm\scriptscriptstyle L}-1}B_{{\rm {\rm\scriptscriptstyle L}}}\,\big]$,
$\rho_{{u}}$ be such that ${\mathcal{{U}}}\subseteq\mathcal{B}_{\rho_{{u}}}(0)$.
%
%
We now introduce the following technical assumption.
\begin{assumption}\label{Assump:term_constr}\hfill
\begin{enumerate}
\item $\|A_{\rm\scriptscriptstyle L}^{N_{\rm\scriptscriptstyle L}}\|<1$;
\item for each $i=1,...,M$, letting ${\mathcal{R}_{i}}(N_{\rm\scriptscriptstyle L})=\begin{bmatrix}(A_{{\rm {\rm\scriptscriptstyle L}}}^{ii})^{N_{\rm\scriptscriptstyle L}-1}B_{{\rm {\rm\scriptscriptstyle L}}}^{ii}&\dots&B_{{\rm {\rm\scriptscriptstyle L}}}^{ii}\end{bmatrix}$ be the reachability matrix in $N_{\rm\scriptscriptstyle L}$ steps associated to $(A_{{\rm {\rm\scriptscriptstyle L}}}^{ii},B_{{\rm {\rm\scriptscriptstyle L}}}^{ii})$,
matrix
\[
{\mathcal{H}}_{i}(N_{\rm\scriptscriptstyle L})=\beta_{i}{\mathcal{R}}_{i}(N_{\rm\scriptscriptstyle L})\in{\mathbb{R}}^{\bar{n}_{i}\times Nm_{i}}
\]
is full-rank with minimum singular value $\underline{\sigma}_{\mathcal{H}_{i}(N_{\rm\scriptscriptstyle L})}>0$;\label{Assump:term_constr_rk}
\item letting $\rho_{\bar{u}}$ and $\rho_{\delta\hat{u}_{i}}$ be such that
${\mathcal{\bar{U}}}\subseteq\mathcal{B}_{\rho_{\bar{u}}}(0)$
and ${\Delta\hat{\mathcal{U}}_{i}}\supseteq\mathcal{B}_{\rho_{\delta\hat{u}_{i}}}(0)$,
respectively, for any $i=1,...,M$ it holds that
\begin{equation}\label{eq:rho_delta_u}
  \rho_{\delta\hat{u}_{i}}>\frac{\kappa(N_{\rm\scriptscriptstyle L})\rho_{\bar{u}}}{\sqrt{N_{\rm\scriptscriptstyle L}}\underline{\sigma}_{\mathcal{H}_{i}(N_{\rm\scriptscriptstyle L})}}
\end{equation}\label{Assump:term_constr_kappa}
\item for each $i=1,\dots,M$
\begin{equation}
\chi_{i}(kN_{\rm\scriptscriptstyle L})=\frac{\sqrt{N_{\rm\scriptscriptstyle L}}\varrho_{u}\|{\mathcal{R}(N_{\rm\scriptscriptstyle L})}\|\|{\mathcal{A}}(N_{\rm\scriptscriptstyle L})\|}{(1-\|A_{\rm\scriptscriptstyle L}^{N_{\rm\scriptscriptstyle L}}\|)(\sqrt{N_{\rm\scriptscriptstyle L}}\underline{\sigma}_{\mathcal{H}_{i}(N_{\rm\scriptscriptstyle L})}\rho_{\delta \hat{u}_{i}}-\kappa(N_{\rm\scriptscriptstyle L})\rho_{\bar{u}})}\leq1\label{Eqn:chidef}
\end{equation}
\item Defining $\Delta\bar{\mathcal{U}}_i=\Delta\mathcal{U}_i(N_{\rm\scriptscriptstyle L}-1)$, and $\Delta\mathcal{U}_i(j)=\Delta\hat{\mathcal{U}}_i\oplus \mathcal{B}_{\rho_{\Delta{u}_i}(j)}(0)$ where $\rho_{\Delta{u}_i}(j)=\sum_{r=2}^j\|K_i I_{si} F_{\rm\scriptscriptstyle L}^{j-r}(A_{\rm\scriptscriptstyle L}-A_{\rm\scriptscriptstyle L}^D)\|\\ \rho_{\delta\hat{x}}(r-1)$ for all $j=2,\dots,N_{\rm\scriptscriptstyle L}$, $\rho_{\Delta{u}_i}(j)=0$ for $j=0,1$ we require that
    \begin{equation}\bar{\mathcal{U}}\oplus(\prod_{i=1}^M \Delta \bar{\mathcal{U}}_i)\subseteq \mathcal{U}\label{eq:bound u_def}\end{equation}\label{Assumption_bound_u_def} \hfill$\square$
\end{enumerate}
\end{assumption}

It is now possible to specify the size of the uncertainty set $\mathcal{W}$ to be considered in the high level design. Specifically, let
\begin{equation}
\label{eq:W_def}
\mathcal{W}=\mathcal{B}_{\rho_w}(0)
\end{equation}
where $\rho_w=\sum_{j=2}^{N_{\rm\scriptscriptstyle L}}\|\beta F_{\rm\scriptscriptstyle L}^{N_{\rm\scriptscriptstyle L}-j}(A_{\rm\scriptscriptstyle L}-A_{\rm\scriptscriptstyle L}^D)\|\rho_{\delta\hat{x}}(j-1)$, $\rho_{\delta\hat{x}}(j)= \sqrt{\sum_{i=1}^M\rho^2_{\delta\hat{x}_i}(j)}$,
\begin{equation}\rho_{\delta\hat{x}_i}(j)=\rho_{\delta \hat{u}_i}\sum_{r=1}^{j}\|(A_{\rm\scriptscriptstyle L}^{ii})^{j-r}B_{\rm\scriptscriptstyle L}^{ii}\|\label{Eqn:bound_deltaxhat}\end{equation}
and where $A_{\rm\scriptscriptstyle L}^D=$diag$(A_{\rm\scriptscriptstyle L}^{11},\dots,A_{\rm\scriptscriptstyle L}^{MM})$.\\
The following result can be proved.
\begin{theorem}
\label{theorem:overall_feasibility}
Under Assumption \ref{Assump:term_constr}, if $x(0)$ is such that the problem \eqref{Eqn:HLoptimiz_1} is feasible at $k=0$ and, for all $i=1,\dots,M$ $$\|x(0)\|\leq{\frac{(\sqrt{N_{\rm\scriptscriptstyle L}}\underline{\sigma}_{\mathcal{H}_{i}(N_{\rm\scriptscriptstyle L})}\rho_{\delta \hat{u}_{i}}-\kappa(N_{\rm\scriptscriptstyle L})\rho_{\bar{u}})}{\|{\mathcal{A}}(N_{\rm\scriptscriptstyle L})\|}}:=\lambda_{i}(N_{\rm\scriptscriptstyle L})$$
then\\
(i) $\bar{w}(k)\in\mathcal{W}$ and problems \eqref{Eqn:HLoptimiz_1} and \eqref{Eqn:LLoptimiz_1} are feasible for all $k\geq 0$;\\
(ii) for all $h\geq 0$
\begin{equation}u(h)\in\mathcal{U}\label{eq:bound u}\end{equation}
(iii) the state of the slow time-scale reduced model $\bar{\Sigma}^{[N_{\rm\scriptscriptstyle L}]}$ enjoys robust convergence properties, i.e.,
$$\bar{x}^{[N_{\rm\scriptscriptstyle L}]}(k)\rightarrow \mathcal{Z}\text{ as }k\rightarrow+\infty$$
(iv)  the state of the large scale model $\Sigma$ enjoys robust convergence properties, i.e., for a computable positive constant $\rho_x$
$$x(k N_{\rm\scriptscriptstyle L})\rightarrow \bigoplus_{h=0}^\infty (F_{\rm\scriptscriptstyle L}^{[N_{\rm\scriptscriptstyle L}]})^h \mathcal{B}_{\rho_x}(0)$$
\hfill$\square$
\end{theorem}
Theorem \ref{theorem:overall_feasibility} establishes two important facts. First, it shows that, if the initial state lies in a suitable set (and Assumption \ref{Assump:term_constr} holds), the joint feasibility properties of the two control layers can be guaranteed in a recursive fashion. Secondly, it ensures convergence of the state of the small-scale slow system considered by the higher control layer to a set.\\
Regarding the main technical Assumption \ref{Assump:term_constr}, note that it involves quantities, that are all functions of the number of steps $N_{\rm\scriptscriptstyle L}$. It is worth now analysing their dependence upon it. Indeed, the following facts can be proved.

\begin{itemize}
\item In view of Assumption~\ref{Assump:3}.\ref{Assump:3_1}, $\beta_{i}$ are full rank and, in view of Assumption~\ref{eq:Assump_LL}.\ref{eq:Assump_LL_Reach}, the pairs $(A_{{\rm {\rm\scriptscriptstyle L}}}^{ii},B_{{\rm {\rm\scriptscriptstyle L}}}^{ii})$
are reachable, so that Assumption~\ref{Assump:term_constr}.\ref{Assump:term_constr_rk}
is fulfilled by taking $N_{\rm\scriptscriptstyle L}$ sufficiently large, i.e., by making the upper layer slower without modifying the rate of the lower layer.
\item In view of Assumptions~\ref{eq:Assump_LL}.\ref{eq:Assump_LL_Schur},~\ref{Assump:3}.\ref{Assump:3_2}, and~\ref{Assump:3}.\ref{Assump:3_0}, $\kappa(N_{\rm\scriptscriptstyle L})=\|\sum_{j=0}^{N_{\rm\scriptscriptstyle L}-1}A_{\rm\scriptscriptstyle H}^{j}B_{\rm\scriptscriptstyle H}-G_{\rm\scriptscriptstyle H}(1)-(\beta\sum_{j=0}^{N_{\rm\scriptscriptstyle L}-1}A_{\rm\scriptscriptstyle L}^{j}B_{\rm\scriptscriptstyle L}-\hat{G}_{\rm\scriptscriptstyle L}(1))\|$ and
$G_{\rm\scriptscriptstyle H}(1)=\sum_{j=0}^{+\infty}A_{\rm\scriptscriptstyle H}^{j}B_{\rm\scriptscriptstyle H}$, $\hat{G}_{\rm\scriptscriptstyle L}(1)=\beta\sum_{j=0}^{+\infty}A_{\rm\scriptscriptstyle L}^{j}B_{\rm\scriptscriptstyle L}$. Therefore
$\kappa(N_{\rm\scriptscriptstyle L})\leq \|\sum_{j=N_{\rm\scriptscriptstyle L}}^{+\infty}A_{\rm\scriptscriptstyle H}^{j}B_{\rm\scriptscriptstyle H}\|+\|\beta\sum_{j=N_{\rm\scriptscriptstyle L}}^{+\infty}A_{\rm\scriptscriptstyle L}^{j}B_{\rm\scriptscriptstyle L}\|\leq \|A_{\rm\scriptscriptstyle H}^{N_{\rm\scriptscriptstyle L}}\| \|G_{\rm\scriptscriptstyle H}(1)\|\\+\|A_{\rm\scriptscriptstyle L}^{N_{\rm\scriptscriptstyle L}}\| \|\beta\|\|G_{\rm\scriptscriptstyle L}(1)\|$,
where ${G}_{{\rm {\rm\scriptscriptstyle L}}}(z)=(zI-A_{{\rm {\rm\scriptscriptstyle L}}})^{-1}B_{{\rm {\rm\scriptscriptstyle L}}}$. Therefore $\kappa(N_{\rm\scriptscriptstyle L})\rightarrow 0$ exponentially as $N_{\rm\scriptscriptstyle L}\rightarrow +\infty$. This shows that also Assumption~\ref{Assump:term_constr}.\ref{Assump:term_constr_kappa} can be fulfilled by taking $N_{\rm\scriptscriptstyle L}$ sufficiently large.
\item Equivalently to Proposition 2.3 in Picasso et al., for any $i=1,...,M$ it can be proved that
\begin{equation}
\lim_{N_{\rm\scriptscriptstyle L}\to+\infty}\lambda_{i}(N_{\rm\scriptscriptstyle L})=+\infty,\lim_{N_{\rm\scriptscriptstyle L}\to+\infty}\|A_{{\rm {\rm\scriptscriptstyle L}}}^{N_{\rm\scriptscriptstyle L}}\|=0,\lim_{N_{\rm\scriptscriptstyle L}\to+\infty}\chi_{i}(N_{\rm\scriptscriptstyle L})=0\label{Eqn:thirdproperty}
\end{equation}
\item The above considerations also show that, by setting a sufficiently large low-level prediction horizon $N_{\rm\scriptscriptstyle L}$, it is always possible to allow for arbitrarily small input constraint sets $\Delta\hat{\mathcal{U}}_i$. This, in turn, allows to obtain an arbitrarily small high-level disturbance set $\mathcal{W}$ and, in turn, a small robust positively invariant set $\mathcal{Z}$ which, eventually, allows to define the input sets in such a way that \eqref{eq:bound u_def} can be verified.
\end{itemize}
The implementation of the multilayer algorithm described in the previous section requires a number of off-line computations here listed for the reader's convenience.
\begin{itemize}
  \item  design of $A_{\rm\scriptscriptstyle H}$, $B_{\rm\scriptscriptstyle H}$, and $\beta_i$, $i=1,...,M$, such that Assumption \eqref{Assump:3} is satisfied;
  \item design of $\bar{K}_{\rm\scriptscriptstyle H}$ such that both $F_{\rm\scriptscriptstyle H}=A_{{\rm {\rm\scriptscriptstyle H}}}^{N_{\rm\scriptscriptstyle L}}+B_{{\rm {\rm\scriptscriptstyle H}}}^{{[N_{\rm\scriptscriptstyle L}]}}\bar{K}_{\rm\scriptscriptstyle H}$ and
$F_{\rm\scriptscriptstyle L}^{[N_{\rm\scriptscriptstyle L}]}=A_{\rm\scriptscriptstyle L}^{N_{\rm\scriptscriptstyle L}}+B_{\rm\scriptscriptstyle L}^{[N_{\rm\scriptscriptstyle L}]}\bar{K}_{\rm\scriptscriptstyle H}\beta$ are Schur stable;
  \item design of $K=diag(K_1,\dots,K_M)$ such that $F_{\rm\scriptscriptstyle L}=A_{\rm\scriptscriptstyle L}+B_{\rm\scriptscriptstyle L}K$ is Schur stable;
  \item computation of $\rho_{\delta \hat{u}_i}$, $\rho_{\bar{u}_i}$ (see the procedure proposed in Appendix \ref{app:rhos}) and of the sets ${\bar{\mathcal{U}}}_{i}$, $\Delta\hat{\mathcal{U}}_i$;
  \item computation of $\mathcal{W}$  according to \eqref{eq:W_def} and \eqref{Eqn:bound_deltaxhat};
  \item computation of $\bar{\mathcal{X}}_{\rm\scriptscriptstyle F}$, $\mathcal{Z}$, see \cite{RawlingsBook}, and $P_{\rm\scriptscriptstyle H}$ with \eqref{eqn:HL_Lyap}.
\end{itemize}
\section{Simulation example}

\begin{figure}[H]
\center
\includegraphics[width=8cm,height=4cm]{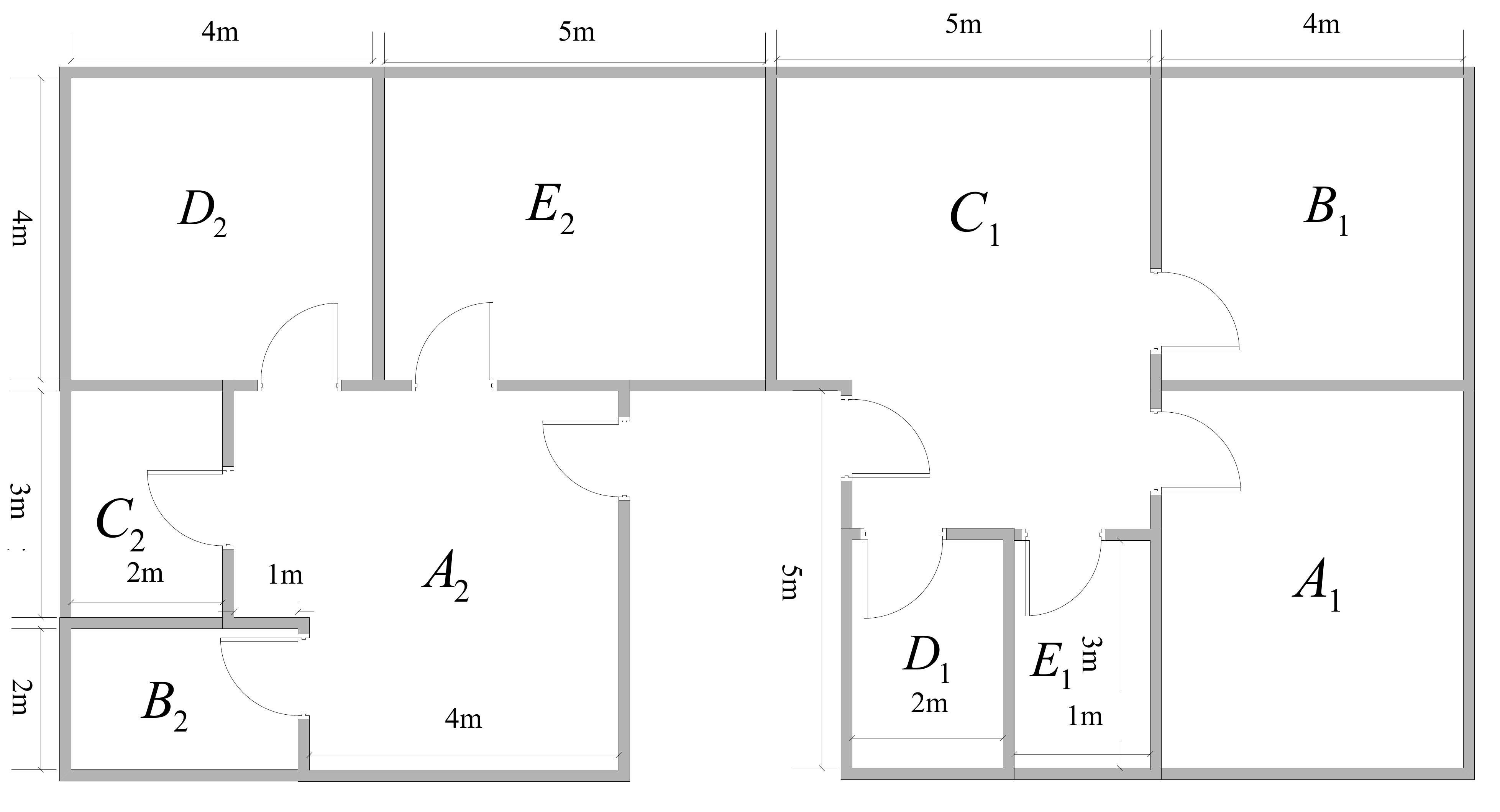}
\caption{Schematic representation of a building with two apartments}
\label{fig:twoflat}
\end{figure}

Consider the problem of regulating temperatures of two apartments depicted in Figure~\ref{fig:twoflat}. The first apartment is constituted by rooms $A_1$, $B_1$, $C_1$, $D_1$ and $E_1$,  while the second one by $A_2$, $B_2$, $C_2$, $D_2$ and $E_2$. Each apartment is equipped with a radiator supplying heats $q_i$, $i=1,\,2$. Heat exchange coefficient between  neighbouring rooms of different apartment, i.e., $E_2$ and $C_1$, is $k_1^t=1$ W/m$^2$K, the one between adjacent rooms inside each apartment is $k_2^t=2.5$ W/m$^2$K, and the one between the rooms and the external environment is $k_e^t=0.5$ W/m$^2$K. The external temperature is $T_E=0 ^{\circ}\mathrm{C}$ and, for simplicity, we neglect solar radiation. Furthermore, the height of the walls is $H=4$ m.  Air density and heat capacity are $\rho=1.225$ kg/m$^3$ and $c=1005$ J/kgK, respectively.
The overall model is made by dynamic energy balance equations of each room. The variables $q_1,\,q_2$ are expressed in Watts, while all the temperature variables are expressed in $^{\circ} \mathrm{C}$. The considered equilibrium point is: $\bar{q}=(\bar{q}_1,\bar{q}_2)=(354.2,320.8)$, with $\bar{T}=(\bar{T}_{A_1},\bar{T}_{B_1},\bar{T}_{C_1},\bar{T}_{D_1},\bar{T}_{E_1},\bar{T}_{A_2},\bar{T}_{B_2},\bar{T}_{C_2},\bar{T}_{D_2},\bar{T}_{E_2})=
(19.6, 20.3,20.2,\\21.7,18.2,17.2,21.2,21.7,19.6,19.4)$. Let $\delta T_{j_i}=T_{j_i}-\bar{T}_{j_i} $ and $\delta q_i=q_i-\bar{q}_i$, for $j=A,\,B,\,C,\,D$ and $i=1,\,2$. In this way, $x_i=(\delta T_{A_i},\, \delta T_{B_i},\, \delta T_{C_i},\, \delta T_{D_i},\,\delta T_{E_i})$ and $u_i=\delta q_i$  are the state and input variables of the $i$-th subsystem , i.e., $n_{i}=5$ and $m_{i}=1$, with $i=1,\,2$.\\
The control variables are limited, i.e., $-50\leqslant u_{1},u_{2}\leqslant50$.

The two subsystems' continuous-time models have been sampled using
the algorithm described in \cite{farina2013block} with
$\Delta t=90\rm s$ to obtain
their discrete-time counterpart in the fast time scales. The eigenvalues of the first subsystem  are $\{0.73,\,0.97,\,0.9,\,
0.85,\,0.88\}$, and the eigenvalues of the second one are $\{0.97,\,0.76,\,0.82,\,0.91,\,0.87\}$. Then the procedure described in the Section (1) has been used to compute the discrete-time reduced
order model, with $A_{\rm\scriptscriptstyle H}=\rm diag(0.97,0.97)$, i.e., $\bar{n}=2$, as well as the transformation matrices $\beta_{1}=\begin{bmatrix}0&0&-1&0&0\end{bmatrix}$ and $\beta_{2}=\begin{bmatrix}0&0&0&0&-1\end{bmatrix}$. The plant model in the slow time scale has been constructed with $N_{\rm\scriptscriptstyle L}=20$.

Tube-based Robust MPC has been designed at the high level according to the algorithm
described in \cite{Mayne2005219} with prediction horizon, $N_{\rm\scriptscriptstyle H}=10$,
state and input penalty, $Q_{\rm\scriptscriptstyle H}=I_{\bar{n}}$ and $R_{\rm\scriptscriptstyle H}=0.1I_{\bar{m}}$.

At the low level, the finite-horizon optimization algorithms described
in \eqref{Eqn:LLoptimiz_1} have been implemented with $Q_{1}=I_{n_{1}}$, $Q_{2}=I_{n_{2}}$,
$R_{1}=R_{2}=10$.

\begin{figure}[ht]
\center
\includegraphics[width=0.9\columnwidth]{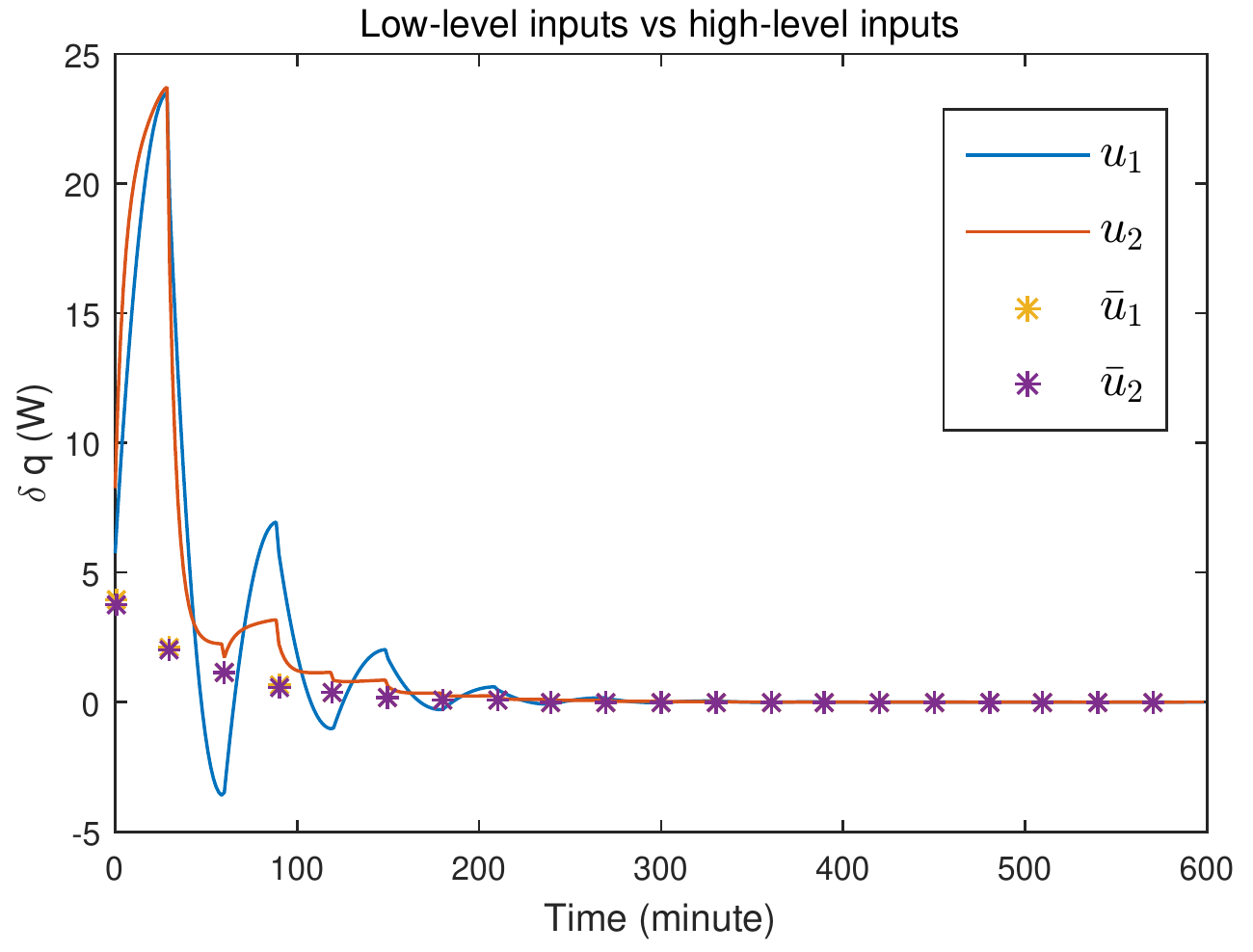}
\caption{Inputs of the two apartments at the high and low levels.}
\label{fig:INP}
\end{figure}

\begin{figure}[ht]
\center
\includegraphics[width=0.9\columnwidth]{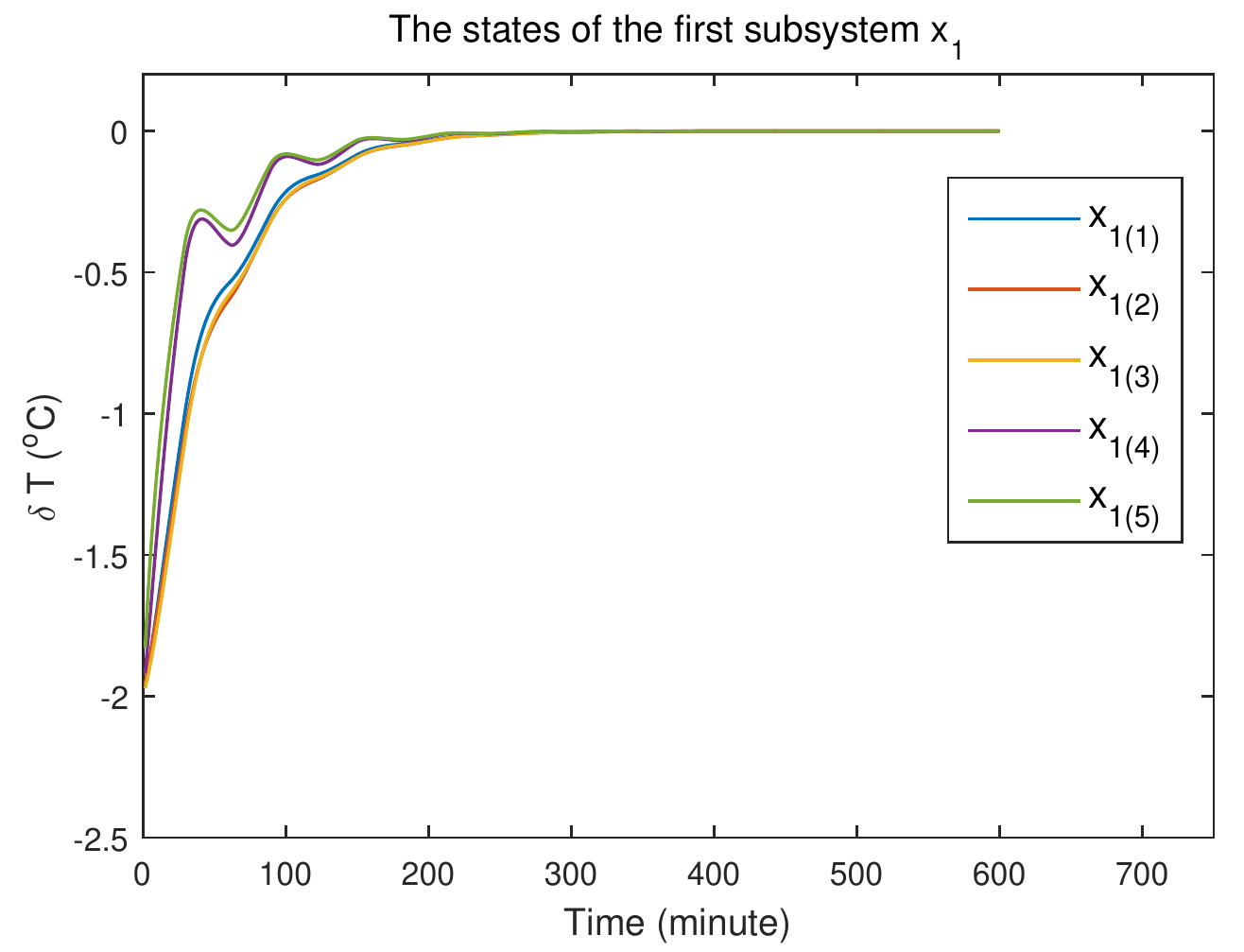}
\caption{Room temperatures in the first apartment.}
\label{fig:ST1}
\end{figure}

\begin{figure}[ht]
	\center
	\includegraphics[width=0.9\columnwidth]{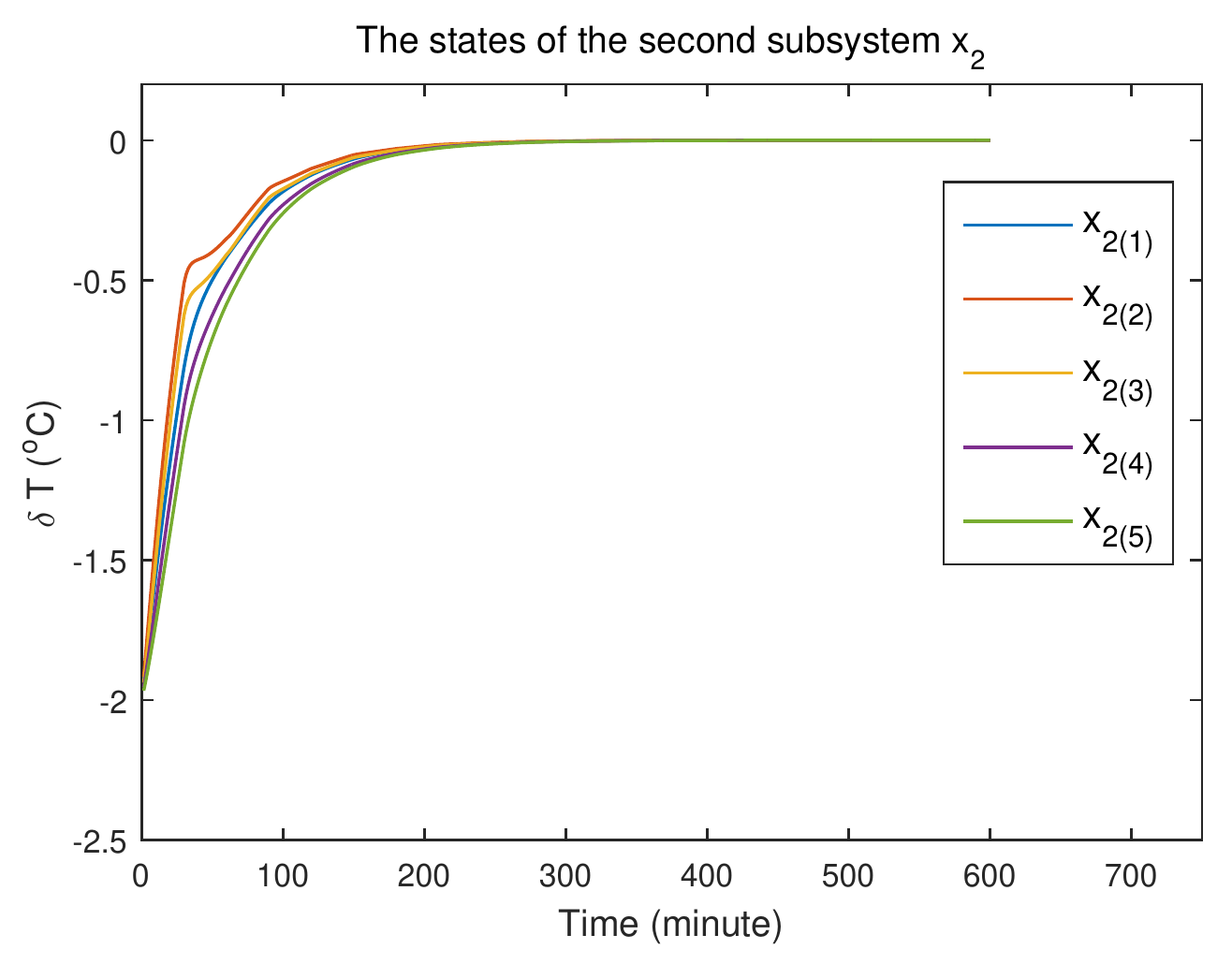}
	\caption{Room temperatures in the second apartment.}
	\label{fig:ST2}
\end{figure}

The hierarchical control scheme has been simulated starting from $x(0)=(x_1(0),x_2(0))=(-2,\dots,-2)$
and $\bar{x}(0)=\beta x(0)$. The transients of the state and control
variables are reported in Figures~\ref{fig:INP}-~\ref{fig:ST2}. These results show the effectiveness
of the proposed algorithm.

\section{Conclusions}
A two layer control scheme for systems made by interconnected subsystems has been presented. Its performance has been tested in simulation, and its properties of recursive feasibility and convergence to a set have been established. Current research is focusing on the extension of the analysis to guarantee convergence to the origin and to deal with tracking problems, as well as the application of the proposed approach to other systems with large dimensions.
%
%
%
%
%
\renewcommand \thesection {APPENDIX}
\section{}
\renewcommand\thesection{A}
\renewcommand \thesubsection {\thesection.\arabic{subsection}}

\subsection{Computation of the input constraint sets}
\label{app:rhos}
In the scheme proposed in this paper, the dimensions of the input constraint sets $\bar{\mathcal{U}}_i$ and $\Delta\hat{\mathcal{U}}_i$ are key tuning knobs. They must be tuned in order to satisfy, at the same time, the inequalities~\eqref{eq:rho_delta_u}, for all $i=1,\dots,M$ and~\eqref{eq:bound     u_def}. To address the design issue, in this appendix we propose a simple and lightweight algorithm based on a linear program. As a simplifying assumption, we set $\Delta\hat{\mathcal{U}}_i=\mathcal{B}_{\rho_{\delta \hat{u}_i}}(0)$ and $\bar{\mathcal{U}}_i=\mathcal{B}_{\rho_{\bar{u}_i}}(0)$. Under this assumption, the tuning knobs are the vectors $\overrightarrow{\rho}_{\delta\hat{u}}=(\rho_{\delta\hat{u}_1},\dots,\rho_{\delta\hat{u}_M})$ and $\overrightarrow{\rho}_{\bar{u}}=(\rho_{\bar{u}_1},\dots,\rho_{\bar{u}_M})$. Note that, in case of need, such assumption can be relaxed, at the price of a slightly different definition of the inequalities below.

First consider inequality~\eqref{eq:rho_delta_u}, to be verified for all $i=1,\dots,M$. Here the constant $\rho_{\bar{u}}$ appears, defined in such a way that $\bar{\mathcal{U}}=\prod_{i=1}^M\mathcal{B}_{\rho_{\bar{u}_i}}(0)\subseteq \mathcal{B}_{\rho_{\bar{u}}}(0)$. We can define, for example, $\rho_{\bar{u}}=\sqrt{\sum_{i=1}^M \rho_{\bar{u}_i}^2}\leq \sum_{i=1}^M \rho_{\bar{u}_i}$. Therefore, to fulfill~\eqref{eq:rho_delta_u} it is sufficient to verify the following matrix inequality
\begin{equation}\label{eq:rho_delta_u_2}
  \overrightarrow{\rho}_{\delta\hat{u}}>\frac{\kappa(N_{\rm\scriptscriptstyle L})}{\sqrt{N_{\rm\scriptscriptstyle L}}}\rm{diag}(\frac{1}{\underline{\sigma}_{\mathcal{H}_{1}(N_{\rm\scriptscriptstyle L})}},\dots,\frac{1}{\underline{\sigma}_{\mathcal{H}_{M}(N_{\rm\scriptscriptstyle L})}})\mathds{1}_{M\times M}\overrightarrow{\rho}_{\bar{u}}
\end{equation}
where $\mathds{1}_{M\times M}$ is the $M\times M$ matrix whose entries are all equal to~$1$.
The second main inclusion to be fulfilled is~\eqref{eq:bound     u_def}, which is verified if, for all $i=1,\dots,M$,
 \begin{equation}\label{eq:bound     u_i}
   \Delta \bar{\mathcal{U}}_i\oplus\bar{\mathcal{U}}_i \subseteq \mathcal{U}_i
 \end{equation}
By definition, $\Delta \bar{\mathcal{U}}_i=\Delta\hat{\mathcal{U}}_i\oplus\mathcal{B}_{\rho_{\Delta u_i(N_{\rm\scriptscriptstyle L}-1)}}(0)$, where
\begin{equation}\label{eq:lambda_ij}
  \begin{array}{cll}
 & \rho_{\Delta{u}_i}(N_{\rm\scriptscriptstyle L}-1)=\\
 &=\sum_{r=2}^{N_{\rm\scriptscriptstyle L}-1}\|K_i I_{si} F_{\rm\scriptscriptstyle L}^{N_{\rm\scriptscriptstyle L}-r-1}(A_{\rm\scriptscriptstyle L}-A_{\rm\scriptscriptstyle L}^D)\|\sqrt{\sum_{j=1}^{M}\rho^2_{\delta \hat{x}_j}(r-1)}\\
&\leq\sum_{r=2}^{N_{\rm\scriptscriptstyle L}-1}\|K_i I_{si} F_{\rm\scriptscriptstyle L}^{N_{\rm\scriptscriptstyle L}-r-1}(A_{\rm\scriptscriptstyle L}-A_{\rm\scriptscriptstyle L}^D)\|\sum_{j=1}^{M}\rho_{\delta \hat{x}_j}(r-1) \\
 & =\sum_{r=2}^{N_{\rm\scriptscriptstyle L}-1}\|K_i I_{si} F_{\rm\scriptscriptstyle L}^{N_{\rm\scriptscriptstyle L}-r-1}(A_{\rm\scriptscriptstyle L}-A_{\rm\scriptscriptstyle L}^D)\|\sum_{j=1}^{M}\sum_{k=1}^{r-1}\|(A_{\rm\scriptscriptstyle L}^{jj})^{r-1-k}B_{\rm\scriptscriptstyle L}^{jj}\|\rho_{\delta \hat{u}_j}\\
 &=\sum_{j=1}^{M}\lambda_{ij}\rho_{\delta \hat{u}_j}
\end{array}
\end{equation}
where $\lambda_{ij}=\sum_{r=2}^{N_{\rm\scriptscriptstyle L}-1}\|K_i I_{si} F_{\rm\scriptscriptstyle L}^{N_{\rm\scriptscriptstyle L}-r-1}(A_{\rm\scriptscriptstyle L}-A_{\rm\scriptscriptstyle L}^D)\|\sum_{k=1}^{r-1}\|(A_{\rm\scriptscriptstyle L}^{jj})^{r-1-k}\\B_{\rm\scriptscriptstyle L}^{jj}\|$. This implies that
$\Delta \bar{\mathcal{U}}_i=\mathcal{B}_{\rho_{\delta \hat{u}_i}+\sum_{j=1}^{M}\lambda_{ij}\rho_{\delta \hat{u}_j}}(0)$. Therefore, to verify~\eqref{eq:bound     u_i} it is sufficient to enforce the constraint
\begin{equation}\label{eq:c de u hat}
({\Lambda}+I){\overrightarrow{\rho}_{\delta\hat{u}}}+{\overrightarrow{\rho}_{\bar{u}}}\leq {\overrightarrow{\rho}_{{u}}}
\end{equation}
where ${\Lambda}$ is the $M\times M$ matrix whose entries are $\lambda_{ij}$, $i,j=1,\ldots,M$, while $\overrightarrow{\rho}_{{u}}=(\rho_{u_1},\dots,\rho_{u_M})$, where $\mathcal{B}_{\rho_i}(0)\subseteq \mathcal{U}_i$ for all $i=1,\dots,M$.
Eventually, a suitable choice of $\overrightarrow{\rho}_{\delta\hat{u}}$ and $\overrightarrow{\rho}_{\bar{u}}$ is obtained as the solution to the following linear programming problem:
\begin{equation}
\begin{array}{cll}
\mbox{max} &  & J_{\rho}\\
\overrightarrow{\rho}_{\delta\hat{u}},\overrightarrow{\rho}_{\bar{u}}\\
\mbox{subject to}
 &  & \rm{constraint}\,\, \eqref{eq:rho_delta_u_2} \,\,\rm{and}\,\, \eqref{eq:c de u hat}
\end{array}\label{Eqn:linearoptmiz_u}
\end{equation}
where $J_{\rho}=\gamma_1 \mathds{1}_{1\times M}\overrightarrow{\rho}_{\delta\hat{u}}+\gamma_2 \mathds{1}_{1\times M}\overrightarrow{\rho}_{\bar{u}}$, where $\gamma_1$, $\gamma_2$ are arbitrary positive weighting constants.

\subsection{Proof of Theorem \ref{theorem:overall_feasibility}}
\label{app:proof}
The proof of Theorem \ref{theorem:overall_feasibility} lies on the intermediary results stated below.
\begin{proposition}\label{prop:feasibility1}\hfill{}\\
A) Under Assumption~\ref{Assump:term_constr} and if $\bar{x}^{[N_{\rm\scriptscriptstyle L}]}(k)=\beta x(kN_{\rm\scriptscriptstyle L})$,
then for any initial condition $\hat{x}(kN_{\rm\scriptscriptstyle L})=x(kN_{\rm\scriptscriptstyle L})$ such that, for all $i=1,\dots,M$
\begin{equation}
\|x(kN_{\rm\scriptscriptstyle L})\|\leq \lambda_{i}(N_{\rm\scriptscriptstyle L})\label{Eqn:condition1}
\end{equation}
and for any $\bar{u}^{{[N_{\rm\scriptscriptstyle L}]}}\in{\mathcal{\bar{U}}}$
there exists a feasible sequence\break\\ $\overrightarrow{\delta \hat{u}}_{i}{ (kN_{\rm\scriptscriptstyle L}:kN_{\rm\scriptscriptstyle L}+N_{\rm\scriptscriptstyle L}-1|kN_{\rm\scriptscriptstyle L})}\in{\varDelta\hat{\mathcal{ U}}_{i}}^{N_{\rm\scriptscriptstyle L}}$
such that the terminal constraint~\eqref{eq:terminaln} is satisfied.\\
B) if $x(kN_{\rm\scriptscriptstyle L})$ satisfies condition~\eqref{Eqn:condition1}, $\|A_{\rm\scriptscriptstyle L}^{N_{\rm\scriptscriptstyle L}}\|<1$,
and, for all $i=1,\dots,M$, \eqref{Eqn:chidef} is verified, then recursive feasibility of the terminal constraint~\eqref{eq:terminaln} is guaranteed.
\end{proposition}

\noindent
\emph{Proof of Proposition \ref{prop:feasibility1}}\\\\
A) Consider the constraint~\eqref{eq:terminaln} and first note that,
since $\delta\hat{x}_{i}(kN_{\rm\scriptscriptstyle L})=0$,
\begin{equation}
\beta_{i}\delta\hat{x}_{i}(kN_{\rm\scriptscriptstyle L}+N_{\rm\scriptscriptstyle L})={\mathcal{H}}_{i}(N_{\rm\scriptscriptstyle L})\overrightarrow{\delta \hat{u}}_{i}{ (kN_{\rm\scriptscriptstyle L}:kN_{\rm\scriptscriptstyle L}+N_{\rm\scriptscriptstyle L}-1|kN_{\rm\scriptscriptstyle L})}\label{Eqn:rp1}
\end{equation}
Moreover, in view of~\eqref{Eqn:C HLS model}
\begin{equation}
\bar{x}^{[N_{\rm\scriptscriptstyle L}]}(k+1)=A_{\rm\scriptscriptstyle H}^{N_{\rm\scriptscriptstyle L}}\beta x(kN_{\rm\scriptscriptstyle L})+\sum_{j=1}^{N_{\rm\scriptscriptstyle L}}A_{\rm\scriptscriptstyle H}^{N_{\rm\scriptscriptstyle L}-j}B_{\rm\scriptscriptstyle H}\bar{u}^{[N_{\rm\scriptscriptstyle L}]}(k)\label{Eqn:rp2}
\end{equation}
Analogously, from~\eqref{Eqn:LL sdyn}
written in collective form
\begin{equation}
\beta\hat{x}(kN_{\rm\scriptscriptstyle L}+N_{\rm\scriptscriptstyle L})=\beta A_{\rm\scriptscriptstyle L}^{N_{\rm\scriptscriptstyle L}}x(kN_{\rm\scriptscriptstyle L})+\beta\sum_{j=1}^{N_{\rm\scriptscriptstyle L}}A_{\rm\scriptscriptstyle L}^{N_{\rm\scriptscriptstyle L}-j}B_{\rm\scriptscriptstyle L}\bar{u}^{[N_{\rm\scriptscriptstyle L}]}(k)\label{Eqn:rp3}
\end{equation}
In view of \eqref{Eqn:rp1}, \eqref{Eqn:rp2}, \eqref{Eqn:rp3}, and the definition
${\mathcal{A}}(N_{\rm\scriptscriptstyle L})$, ${\mathcal{B}}(N_{\rm\scriptscriptstyle L})$, and $I_{si}$, the constraint~\eqref{eq:terminaln}
can be written as
\begin{equation}
\begin{array}{l}
{\mathcal{H}}_{i}(N_{\rm\scriptscriptstyle L})\overrightarrow{\delta \hat{u}}_{i}{(kN_{\rm\scriptscriptstyle L}:kN_{\rm\scriptscriptstyle L}+N_{\rm\scriptscriptstyle L}-1|kN_{\rm\scriptscriptstyle L})}\\
\quad=I_{si}[{\mathcal{A}}(N_{\rm\scriptscriptstyle L})x(kN_{\rm\scriptscriptstyle L})+\mathcal{B}(N_{\rm\scriptscriptstyle L})\bar{u}^{[N_{\rm\scriptscriptstyle L}]}(k)]\end{array}\label{Eqn:p4}
\end{equation}
From this expression, the definitions of $\underline{\sigma}_{\mathcal{H}_{i}(N_{\rm\scriptscriptstyle L})}$,
$\rho_{\bar{u}}$, $\rho_{\delta\hat{u}_{i}}$, and in view of \eqref{Eqn:Gkappa1},
it can be concluded that a feasible sequence $\overrightarrow{\delta \hat{u}}_{i}{(kN_{\rm\scriptscriptstyle L}:kN_{\rm\scriptscriptstyle L}+N_{\rm\scriptscriptstyle L}-1|kN_{\rm\scriptscriptstyle L})}$
can be computed provided that
\begin{equation}
\sqrt{N_{\rm\scriptscriptstyle L}}\underline{\sigma}_{\mathcal{H}_{i}(N_{\rm\scriptscriptstyle L})}\rho_{\delta\hat{u}_{i}}\geq\|\mathcal{A}(N_{\rm\scriptscriptstyle L})\|\|x(kN_{\rm\scriptscriptstyle L})\|+\kappa(N_{\rm\scriptscriptstyle L})\rho_{\bar{u}}\label{Eqn:p5}
\end{equation}
from which the result follows.\\\\
B) From \eqref{Eqn:C full model} it holds that
\begin{equation}
x(kN_{\rm\scriptscriptstyle L}+N_{\rm\scriptscriptstyle L})=A_{\rm\scriptscriptstyle L}^{N_{\rm\scriptscriptstyle L}}x(kN_{\rm\scriptscriptstyle L})+{\mathcal{R}(N_{\rm\scriptscriptstyle L})}\overrightarrow{u}(kN_{\rm\scriptscriptstyle L}:kN_{\rm\scriptscriptstyle L}+N_{\rm\scriptscriptstyle L}-1|kN_{\rm\scriptscriptstyle L})\label{Eqn:p6}
\end{equation}
Therefore
\begin{equation}
\|x(kN_{\rm\scriptscriptstyle L}+N_{\rm\scriptscriptstyle L})\|\leq\|A_{\rm\scriptscriptstyle L}^{N_{\rm\scriptscriptstyle L}}\|\|x(kN_{\rm\scriptscriptstyle L})\|+\sqrt{N_{\rm\scriptscriptstyle L}}\|{\mathcal{R}(N_{\rm\scriptscriptstyle L})}\|\varrho_{u}\label{Eqn:p7}
\end{equation}
and, in view of~\eqref{Eqn:condition1}
\begin{equation}
\begin{array}{c}
\|x(kN_{\rm\scriptscriptstyle L}+N_{\rm\scriptscriptstyle L})\|\leq\|A_{\rm\scriptscriptstyle L}^{N_{\rm\scriptscriptstyle L}}\|{\frac{(\sqrt{N_{\rm\scriptscriptstyle L}}\underline{\sigma}_{\mathcal{H}_{i}(N_{\rm\scriptscriptstyle L})}\rho_{\delta \hat{u}_{i}}-\kappa(N_{\rm\scriptscriptstyle L})\rho_{\bar{u}})}{\|{\mathcal{A}}(N_{\rm\scriptscriptstyle L})\|}}\\
+\sqrt{N_{\rm\scriptscriptstyle L}}\|{\mathcal{R}(N_{\rm\scriptscriptstyle L})}\|\varrho_{u}\label{Eqn:p8}
\end{array}\end{equation}
for all $i=1,2,\dots,M$. From this expression and the definition of $\chi_{i}(kN_{\rm\scriptscriptstyle L})$ through~\eqref{Eqn:chidef}
it turns out that
\begin{equation}
\|x(kN_{\rm\scriptscriptstyle L}+N_{\rm\scriptscriptstyle L})\|\leq{\frac{(\sqrt{N_{\rm\scriptscriptstyle L}}\underline{\sigma}_{\mathcal{H}_{i}(N_{\rm\scriptscriptstyle L})}\rho_{\delta \hat{u}_{i}}-\kappa(N_{\rm\scriptscriptstyle L})\rho_{\bar{u}})}{\|{\mathcal{A}}(N_{\rm\scriptscriptstyle L})\|}}\label{Eqn:p9}
\end{equation}
for all $i=1,2,\dots,M$ and the result follows.\hfill$\square$\\\\

\medskip
\begin{proposition}\label{prop:prop2} If Problem \eqref{Eqn:LLoptimiz_1} is feasible, then\\\\
\noindent
A) For all $k\geq 0$
\begin{equation}\|\bar{w}(k)\|\leq \rho_w\label{Eqn:bound_w_2}\end{equation}
\noindent
B) For all $k\geq 0$,
\begin{equation}\delta u_i(kN_{\rm\scriptscriptstyle L}+j)\in\Delta\mathcal{U}_i(j)\label{Eqn:bound_deltaui}\end{equation}
Also it holds that
\begin{equation}\Delta\bar{\mathcal{U}}_i\supseteq \Delta\mathcal{U}_i(j)\label{Eqn:bound_deltaui_monotonicity}\end{equation}
for all $j=0,\dots,N_{\rm\scriptscriptstyle L}-1$.\hfill$\square$
\end{proposition}

\noindent
\emph{Proof of Proposition \ref{prop:prop2}}\\\\
\noindent
A) Defining the collective vectors $\hat{x}=(\hat{x}_1,\dots,\hat{x}_M)$, $\delta x=(\delta x_1,\dots,\delta x_M)$, $\delta \hat{x}=(\delta \hat{x}_1,\dots,\delta \hat{x}_M)$, and $\varepsilon(kN_{\rm\scriptscriptstyle L}+j|kN_{\rm\scriptscriptstyle L})=\delta x(kN_{\rm\scriptscriptstyle L}+j|kN_{\rm\scriptscriptstyle L})-\delta \hat{x}(kN_{\rm\scriptscriptstyle L}+j|kN_{\rm\scriptscriptstyle L})$, we have that $\bar{w}(k)=\beta x(kN_{\rm\scriptscriptstyle L}+N_{\rm\scriptscriptstyle L})- \bar{x}^{[N_{\rm\scriptscriptstyle L}]}(k+1|k)=\beta \hat{x}(kN_{\rm\scriptscriptstyle L}+N_{\rm\scriptscriptstyle L})+\beta \delta x(kN_{\rm\scriptscriptstyle L}+N_{\rm\scriptscriptstyle L})- \bar{x}^{[N_{\rm\scriptscriptstyle L}]}(k+1|k)=(\beta \hat{x}(kN_{\rm\scriptscriptstyle L}+N_{\rm\scriptscriptstyle L})+\beta \delta \hat{x}(kN_{\rm\scriptscriptstyle L}+N_{\rm\scriptscriptstyle L})- \bar{x}^{[N_{\rm\scriptscriptstyle L}]}(k+1|k))+\beta\varepsilon(kN_{\rm\scriptscriptstyle L}+j|kN_{\rm\scriptscriptstyle L}) = \beta\varepsilon(kN_{\rm\scriptscriptstyle L}+j|kN_{\rm\scriptscriptstyle L})$. The latter equality holds in view of the fact that Problem \eqref{Eqn:LLoptimiz_1} is feasible, and therefore equality \eqref{eq:terminaln} is verified.
From \eqref{Eqn:LL ddyn}, \eqref{Eqn:LL dndyn}, \eqref{Eqn:LL dndyn_sf_deltau}, we collectively have that
\begin{equation}\begin{array}{c}\varepsilon(kN_{\rm\scriptscriptstyle L}+j+1|kN_{\rm\scriptscriptstyle L})=F_{\rm\scriptscriptstyle L} \varepsilon(kN_{\rm\scriptscriptstyle L}+j|kN_{\rm\scriptscriptstyle L})\\
+ (A_{\rm\scriptscriptstyle L}-A_{\rm\scriptscriptstyle L}^D)\delta\hat{x}(kN_{\rm\scriptscriptstyle L}+j|kN_{\rm\scriptscriptstyle L})\end{array}\label{Eqn:epsilon_evolution}\end{equation}
In view of the fact that $\varepsilon(kN_{\rm\scriptscriptstyle L}|kN_{\rm\scriptscriptstyle L})=\delta \hat{x}(kN_{\rm\scriptscriptstyle L}|kN_{\rm\scriptscriptstyle L})=0$, then
$\bar{w}(k)=\beta \sum_{j=2}^{N_{\rm\scriptscriptstyle L}} F_{\rm\scriptscriptstyle L}^{N_{\rm\scriptscriptstyle L}-j}(A_{\rm\scriptscriptstyle L}-A_{\rm\scriptscriptstyle L}^D)\delta\hat{x}(kN_{\rm\scriptscriptstyle L}+j-1|kN_{\rm\scriptscriptstyle L})$. From this it follows that
\begin{equation}\|\bar{w}(k)\|\leq \sum_{j=2}^{N_{\rm\scriptscriptstyle L}} \|\beta F_{\rm\scriptscriptstyle L}^{N_{\rm\scriptscriptstyle L}-j}(A_{\rm\scriptscriptstyle L}-A_{\rm\scriptscriptstyle L}^D)\| \|\delta\hat{x}(kN_{\rm\scriptscriptstyle L}+j-1|kN_{\rm\scriptscriptstyle L})\|\label{Eqn:bound_w_1}\end{equation}
Since $\delta \hat{u}_{i}$ are bounded for all $i=1,\dots,M$, i.e., scalar $\rho_{\delta \hat{u}_i}$ are defined such that $\delta\hat{u}_{i}\in\mathcal{B}_{\rho_{\delta \hat{u}_i}}(0)$. In view of this, we compute that
$\|\delta\hat{x}(kN_{\rm\scriptscriptstyle L}+j|kN_{\rm\scriptscriptstyle L})\|\leq \rho_{\delta\hat{x}}(j)$, where $\rho_{\delta\hat{x}_i}(j)$ is defined in \eqref{Eqn:bound_deltaxhat}. Therefore, $\delta\hat{x}(kN_{\rm\scriptscriptstyle L}+j|kN_{\rm\scriptscriptstyle L})$ are bounded, for all $j=1,\dots,N_{\rm\scriptscriptstyle L}-1$ and more specifically we get that $\|\delta\hat{x}(kN_{\rm\scriptscriptstyle L}+j|kN_{\rm\scriptscriptstyle L})\|\leq \sqrt{\sum_{i=1}^M\rho^2_{\delta\hat{x}_i}(j)}$.
Therefore one has \eqref{Eqn:bound_w_2} for all $k\geq 0$.\\\\
B) From \eqref{Eqn:epsilon_evolution} we have that $\varepsilon(kN_{\rm\scriptscriptstyle L}+j|kN_{\rm\scriptscriptstyle L})=\sum_{r=2}^j F_{\rm\scriptscriptstyle L}^{j-r}(A_{\rm\scriptscriptstyle L}-A_{\rm\scriptscriptstyle L}^D)\delta\hat{x}(kN_{\rm\scriptscriptstyle L}+r-1|kN_{\rm\scriptscriptstyle L})$ and therefore
$\delta u_i(kN_{\rm\scriptscriptstyle L}+j)-\delta \hat{u}_i(kN_{\rm\scriptscriptstyle L}+j|kN_{\rm\scriptscriptstyle L})=K_i \varepsilon_i(kN_{\rm\scriptscriptstyle L}+j|kN_{\rm\scriptscriptstyle L})=K_i I_{si}\sum_{r=2}^j F_{\rm\scriptscriptstyle L}^{j-r}(A_{\rm\scriptscriptstyle L}-A_{\rm\scriptscriptstyle L}^D)\delta\hat{x}(kN_{\rm\scriptscriptstyle L}+r-1|kN_{\rm\scriptscriptstyle L})$. From this it follows that
$\delta u_i(kN_{\rm\scriptscriptstyle L}+j)\in\delta \hat{u}_i(kN_{\rm\scriptscriptstyle L}+j|kN_{\rm\scriptscriptstyle L})\oplus \mathcal{B}_{\rho_{\Delta{u}_i}(j)}(0)$ and therefore $\delta u_i(kN_{\rm\scriptscriptstyle L}+j)\in\Delta\mathcal{U}_i(j)$. In view of the monotonicity property $\rho_{\Delta{u}_i}(j+1)\geq \rho_{\Delta{u}_i}(j)$ for all $j$, it holds that $\mathcal{B}_{\rho_{\Delta{u}_i}(j+1)}(0)\supseteq \mathcal{B}_{\rho_{\Delta{u}_i}(j)}(0)$, which implies \eqref{Eqn:bound_deltaui_monotonicity}.\hfill$\square$\\

\noindent
\emph{Proof of Theorem \ref{theorem:overall_feasibility}}\\\\
(i) If $\|x(0)\|\leq \lambda_{i}(N_{\rm\scriptscriptstyle L})$ and recalling that Assumption~\ref{Assump:term_constr} holds, from Proposition~\ref{prop:feasibility1}, recursive feasibility of the optimization problems~\eqref{Eqn:LLoptimiz_1} is guaranteed, i.e.,  that there exists, for all $k\geq 0$, a feasible sequence $\overrightarrow{\delta \hat{u}}_{i}{ (kN_{\rm\scriptscriptstyle L}:kN_{\rm\scriptscriptstyle L}+N_{\rm\scriptscriptstyle L}-1|kN_{\rm\scriptscriptstyle L})}\in{\varDelta\hat{\mathcal{ U}}_{i}}^{N_{\rm\scriptscriptstyle L}}$
such that the terminal constraint~\eqref{eq:terminaln} is satisfied.\\
Also, from Proposition \ref{prop:prop2}.A, it is proved that $\bar{w}(k)\in\mathcal{W}$ for all $k\geq 0$, which allows to apply the recursive feasibility arguments of \cite{Mayne2005219}, proving that also \eqref{Eqn:HLoptimiz_1} enjoys recursive feasibility properties.\\
(ii) It is now possible to conclude that, in view of the feasibility of \eqref{Eqn:HLoptimiz_1}, $\bar{u}^{[N_{\rm\scriptscriptstyle L}]}(k)\in\bar{\mathcal{U}}$; also, from Proposition~\ref{prop:prop2}.B it follows that $\delta u_{i}(kN_{\rm\scriptscriptstyle L}+j)\in \Delta\bar{\mathcal{U}}_i$ for all $k\geq 0$, $j=0,\dots, N_{\rm\scriptscriptstyle L}-1$, and $i=1,\dots,M$. From this, under \eqref{eq:bound u_def}, the inclusion~\eqref{eq:bound u} can also be proved.\\
(iii) We apply the results in \cite{Mayne2005219}, which guarantee robust convergence properties. In other words, we have that $\bar{x}^{[N_{\rm\scriptscriptstyle L}],o}(k)\rightarrow 0$ as $k\rightarrow +\infty$, and that $\bar{x}^{[N_{\rm\scriptscriptstyle L}]}(k)$ is asymptotically driven to lie in the robust positively invariant set $\mathcal{Z}$.\\
(iv) To show robust convergence of the global system state, from \eqref{Eqn:C full model} we obtain that
\begin{align}x((k+1)N_{\rm\scriptscriptstyle L})=A_{\rm\scriptscriptstyle L}^{N_{\rm\scriptscriptstyle L}}x(kN_{\rm\scriptscriptstyle L})+B_{\rm\scriptscriptstyle L}^{[N_{\rm\scriptscriptstyle L}]}\bar{u}^{[N_{\rm\scriptscriptstyle L}]}(k)\nonumber\\
+\sum_{h=0}^{N_{\rm\scriptscriptstyle L}-1}A_{\rm\scriptscriptstyle L}^hB_{\rm\scriptscriptstyle L}\delta u((k+1)N_{\rm\scriptscriptstyle L}-h-1)\label{eq:evol_x_contr01}\end{align}
Denoting $\mathcal{B}_{L,N_{\rm\scriptscriptstyle L}}^C=\begin{bmatrix}A_{\rm\scriptscriptstyle L}^{N-1}B_{\rm\scriptscriptstyle L}&\dots &B_{\rm\scriptscriptstyle L}\end{bmatrix}$, we obtain that
$$\sum_{h=0}^{N_{\rm\scriptscriptstyle L}-1}A_{\rm\scriptscriptstyle L}^hB_{\rm\scriptscriptstyle L}\delta u((k+1)N_{\rm\scriptscriptstyle L}-h-1)=\mathcal{B}_{L,N_{\rm\scriptscriptstyle L}}^C\overrightarrow{\delta {u}} {(kN_{\rm\scriptscriptstyle L}:kN_{\rm\scriptscriptstyle L}+N_{\rm\scriptscriptstyle L}-1)}$$
Also, recall that
$\overrightarrow{\delta {u}} {(kN_{\rm\scriptscriptstyle L}:kN_{\rm\scriptscriptstyle L}+N_{\rm\scriptscriptstyle L}-1)}=\overrightarrow{\delta \hat{u}} {(kN_{\rm\scriptscriptstyle L}:kN_{\rm\scriptscriptstyle L}+N_{\rm\scriptscriptstyle L}-1)}+{\rm{diag}}(K,\dots,K)\overrightarrow{\varepsilon} {(kN_{\rm\scriptscriptstyle L}:kN_{\rm\scriptscriptstyle L}+N_{\rm\scriptscriptstyle L}-1)}$ and that, by defining
$$\mathcal{F}_{N_{\rm\scriptscriptstyle L}}=\begin{bmatrix}0&0&\dots&0&0&0\\
I&0&\dots&0&0&0\\
F_{\rm\scriptscriptstyle L}&I&\dots&0&0&0\\
\vdots&\vdots&\ddots&\vdots&\vdots&\vdots\\
F_{\rm\scriptscriptstyle L}^{N_{\rm\scriptscriptstyle L}-2}&F_{\rm\scriptscriptstyle L}^{N_{\rm\scriptscriptstyle L}-3}&\dots&I&0&0\end{bmatrix},$$
$$\begin{array}{l}\overrightarrow{\varepsilon} {(kN_{\rm\scriptscriptstyle L}:kN_{\rm\scriptscriptstyle L}+N_{\rm\scriptscriptstyle L}-1)}\\
=\mathcal{F}_{N_{\rm\scriptscriptstyle L}}{\rm{diag}}(A_{\rm\scriptscriptstyle L}^C,\dots,A_{\rm\scriptscriptstyle L}^C)\mathcal{B}_{\rm\scriptscriptstyle L}\overrightarrow{\delta \hat{u}} {(kN_{\rm\scriptscriptstyle L}:kN_{\rm\scriptscriptstyle L}+N_{\rm\scriptscriptstyle L}-1)}\end{array}$$
where $A_{\rm\scriptscriptstyle L}^C=A_{\rm\scriptscriptstyle L}-A_{\rm\scriptscriptstyle L}^D$ and
$$\mathcal{B}_{\rm\scriptscriptstyle L}=\begin{bmatrix}
0&\dots,&0\\
\vdots&\ddots&\vdots\\
(A_{\rm\scriptscriptstyle L}^D)^{N_{\rm\scriptscriptstyle L}-1}B_{\rm\scriptscriptstyle L}&\dots&B_{\rm\scriptscriptstyle L}\end{bmatrix}$$
Recalling that $\bar{x}^{[N_{\rm\scriptscriptstyle L}]}(k)=\beta x(kN_{\rm\scriptscriptstyle L})$ and that
\begin{align}\bar{u}^{[N_{\rm\scriptscriptstyle L}]}(k)=\bar{u}^{[N_{\rm\scriptscriptstyle L}],o}(k)+\bar{K}_{\rm\scriptscriptstyle H} (\bar{x}^{[N_{\rm\scriptscriptstyle L}]}(k)-\bar{x}^{[N_{\rm\scriptscriptstyle L}],o}(k))\label{eq:u^NL_TOT01}\end{align}
we can rewrite \eqref{eq:evol_x_contr01} as
\begin{align}&x((k+1)N_{\rm\scriptscriptstyle L})=(A_{\rm\scriptscriptstyle L}^{N_{\rm\scriptscriptstyle L}}+B_{\rm\scriptscriptstyle L}^{[N_{\rm\scriptscriptstyle L}]}\bar{K}_{\rm\scriptscriptstyle H}\beta)x(kN_{\rm\scriptscriptstyle L})\nonumber\\
&+B_{\rm\scriptscriptstyle L}^{[N_{\rm\scriptscriptstyle L}]}(\bar{u}^{[N_{\rm\scriptscriptstyle L}],o}(k)-\bar{K}_{\rm\scriptscriptstyle H}\bar{x}^{[N_{\rm\scriptscriptstyle L}],o}(k))+\mathcal{B}_{L,N_{\rm\scriptscriptstyle L}}^C\overrightarrow{\delta {u}} {(kN_{\rm\scriptscriptstyle L}:kN_{\rm\scriptscriptstyle L}+N_{\rm\scriptscriptstyle L}-1)}\label{eq:evol_x_contr01bis}\end{align}
Recall that $\bar{x}^{[N_{\rm\scriptscriptstyle L}],o}(k),\,\bar{u}^{[N_{\rm\scriptscriptstyle L}],o}(k)\rightarrow 0$ as $k\rightarrow +\infty$. Also, we compute that
\begin{equation}\begin{array}{c}\overrightarrow{\delta {u}} {(kN_{\rm\scriptscriptstyle L}:kN_{\rm\scriptscriptstyle L}+N_{\rm\scriptscriptstyle L}-1)} =(I+{\rm{diag}}(K_i,\dots,K_i)\mathcal{F}_{N_{\rm\scriptscriptstyle L}}\\
\cdot{\rm{diag}}(A_{\rm\scriptscriptstyle L}^C,\dots,A_{\rm\scriptscriptstyle L}^C)\mathcal{B}_{\rm\scriptscriptstyle L})\overrightarrow{\delta \hat{u}} {(kN_{\rm\scriptscriptstyle L}:kN_{\rm\scriptscriptstyle L}+N_{\rm\scriptscriptstyle L}-1)}\end{array}\label{eq:deltau_tot_new}\end{equation}
Based on this, we define $\kappa_{\delta u}=\|\mathcal{B}_{L,N_{\rm\scriptscriptstyle L}}^C(I+{\rm{diag}}(K_i,\dots,K_i)\mathcal{F}_{N_{\rm\scriptscriptstyle L}}\\
\cdot{\rm{diag}}(A_{\rm\scriptscriptstyle L}^C,\dots,A_{\rm\scriptscriptstyle L}^C)\mathcal{B}_{\rm\scriptscriptstyle L})\|$ and we write \eqref{eq:evol_x_contr01bis} as
\begin{align}x((k+1)N_{\rm\scriptscriptstyle L})=&(A_{\rm\scriptscriptstyle L}^{N_{\rm\scriptscriptstyle L}}+B_{\rm\scriptscriptstyle L}^{[N_{\rm\scriptscriptstyle L}]}\bar{K}_{\rm\scriptscriptstyle H}\beta)x(kN_{\rm\scriptscriptstyle L})\label{eq:evol_x_contr01tris}\\
&+
B_{\rm\scriptscriptstyle L}^{[N_{\rm\scriptscriptstyle L}]}(\bar{u}^{[N_{\rm\scriptscriptstyle L}],o}(k)-\bar{K}_{\rm\scriptscriptstyle H}\bar{x}^{[N_{\rm\scriptscriptstyle L}],o}(k))+w_{\rm\scriptscriptstyle L}(k)\nonumber\end{align}
where $\|w_{\rm\scriptscriptstyle L}(k)\|\leq \kappa_{\delta u} \sqrt{N_{\rm\scriptscriptstyle L}} \max_{h\in\{kN_{\rm\scriptscriptstyle L},\dots,(k+1)N_{\rm\scriptscriptstyle L}-1\}}\|\delta \hat{u}(h)\|$\break
$\leq \kappa_{\delta u} \sqrt{N_{\rm\scriptscriptstyle L}}\sqrt{\sum_{i=1}^M\rho^2_{\delta \hat{u}_i}}$. Therefore, since $F_{\rm\scriptscriptstyle L}^{[N_{\rm\scriptscriptstyle L}]}=A_{\rm\scriptscriptstyle L}^{N_{\rm\scriptscriptstyle L}}+B_{\rm\scriptscriptstyle L}^{[N_{\rm\scriptscriptstyle L}]}\bar{K}_{\rm\scriptscriptstyle H}\beta$ is Schur stable, then the asymptotic result follows, where
$\rho_x=\kappa_{\delta u} \sqrt{N_{\rm\scriptscriptstyle L}} \sqrt{\sum_{i=1}^M\rho^2_{\delta \hat{u}_i}}$.
\hfill{}$\square$%
\bibliographystyle{plain}        
\bibliography{stochasticMPC}
\end{document}